\documentclass[aps,prb,twocolumn,showpacs,superscriptaddress,reprint]{revtex4-1}

\usepackage{mathrsfs}
\usepackage{amsfonts}
\usepackage{amssymb}
\usepackage{amsmath}
\usepackage{graphicx}
\usepackage{sidecap}
\usepackage{physics}
\usepackage{color}
\usepackage[colorlinks=true,citecolor=red,linkcolor=magenta]{hyperref}
\usepackage{subeqnarray}
\usepackage{verbatim}
\usepackage{ulem}
\usepackage{subfigure}
\usepackage{graphicx}
\usepackage{braket}
\usepackage{soul}
\usepackage{float}
\usepackage{array}
\usepackage{tabularx}
\emergencystretch=\maxdimen
\begin{document}

\title{Coherent transfer of singlet-triplet qubit states in an architecture of triple quantum dots}

\author{MengKe Feng}
\affiliation{Division of Physics and Applied Physics, School of Physical and Mathematical Sciences, 21 Nanyang Link, Singapore 637371, Singapore.}
\author{Chang Jian Kwong}\thanks{Author has since moved to Institut f$\ddot{u}$r Experimentalphysik, Heinrich-Heine-Universit$\ddot{a}$t D$\ddot{u}$sseldorf, D$\ddot{u}$sseldorf, Germany}
\affiliation{Centre for Quantum Technologies, National University of Singapore, 3 Science Drive 2, Singapore 117543, Singapore.}
\author{Teck Seng Koh}
\affiliation{Division of Physics and Applied Physics, School of Physical and Mathematical Sciences, 21 Nanyang Link, Singapore 637371, Singapore.}
\author{Leong Chuan Kwek}
\affiliation{Centre for Quantum Technologies, National University of Singapore, 3 Science Drive 2, Singapore 117543, Singapore.}
\affiliation{Institute of Advanced Studies, Nanyang Technological University, 60 Nanyang View, Singapore 639673, Singapore.}
\affiliation{National Institute of Education, Nanyang Technological University, 1 Nanyang Walk, Singapore 637616, Singapore.}
\affiliation{MajuLab, CNRS-UNS-NUS-NTU International Joint Research Unit, UMI 3654, Singapore.}

\begin{abstract}
We propose two schemes to coherently transfer arbitrary quantum states of the two-electron singlet-triplet qubit across a chain of 3 quantum dots. The schemes are based on electrical control over the detuning energy  of the quantum dots. The first is a pulse-gated scheme, requiring dc pulses and engineering of inter- and intra-dot Coulomb energies. The second scheme is based on the adiabatic theorem, requiring time-dependent control of the detuning energy  through avoided crossings at a rate that the system remains in the ground state. We simulate the transfer fidelity using typical experimental parameters for silicon quantum dots. Our results give state transfer fidelities between $94.3\% <\mathcal{F}< 99.5\%$  at sub-ns gate times for the pulse-gated scheme and between $75.4\% <\mathcal{F}< 99.0\%$ at tens of ns for the adiabatic scheme. Taking into account dephasing from charge noise, we obtain state transfer fidelities between $94.0\% <\mathcal{F}< 99.2\%$ for the pulse-gated scheme and between $64.9\% <\mathcal{F}< 93.6\%$ for the adiabatic scheme. 
\end{abstract}

\maketitle

% BACKGROUND AND MOTIVATION 
\section{Introduction}

Spin qubits in semiconductor quantum dots are leading candidates for quantum information processing due to their long coherence times~\cite{Koppens:2006p766, Bluhm:2011p109} and promise of scalability~\cite{Loss:1998p120}. The exchange coupling is the spin-dependent part of the Coulomb interaction between electrons, and is essential in the manipulation of qubit-qubit interaction~\cite{Loss:1998p120}, as well as single-qubit rotations for qubits encoded by two~\cite{Levy:2002p147902, Petta:2010p669} or three~\cite{DiVincenzo:2000p339, Laird:2010p075403} electron spins.

Because of the  architectural and scaling constraints~\cite{Svore:2005p022317} imposed by the short range of the exchange interaction~\cite{Li:2010p085313}, studies of spin qubit architectures invariably involve how quantum information may be transferred from one location to another with high fidelity and experimentally realistic requirements. Existing proposals may be based on moving the electrons themselves~\cite{Taylor:2005p177,flentje_2017,fujita_2017}, or utilising exchange-coupled spin chains that require precise engineering of the exchange interaction~\cite{Christandl:2004p187902}, strong couplings within a ``spin bus''~\cite{Friesen:2007p230503}, or pulse shaping of the tunnel couplings for the single spin qubit~\cite{Greentree:2004p235317} and the triple spin qubit~\cite{Ferraro:2015p075435}. Other proposals based on hybrid systems that transduce spin information into photon modes via a resonant cavity~\cite{Imamoglu:1999p4204, Hu:2012p035314, Petersson:2012p380,Vermersch_2017} introduce new experimental constraints and may be more challenging to realise. 

Despite the successes of the two-electron singlet-triplet (ST) qubit~\cite{Wu:2014p11938, Shulman:2012p202, Maune:2012p344, Bluhm:2011p109, Reilly:2008p817, Foletti:2009p903}, relatively little attention has been given to elucidate techniques for the transfer of quantum information encoded by the ST qubit  without posing additional experimental challenges.

In this paper, we study the coherent transfer of quantum information encoded in the singlet ($S$) and unpolarised triplet ($T_0$) states of the ST qubit across a chain of 3 quantum dots, as sketched in both Fig.~\ref{fig:PulseGated}(a) and~\ref{fig:Adiabatic}(a).  We investigate  the rate and fidelity of the transfer for two schemes that are within reach of current experimental techniques, requiring only control over the detuning energy $\varepsilon$ of individual quantum dots through applied dc and linear voltage pulses. Even though in realistic systems, other parameters, e.g. tunnel coupling, may be cross-coupled to  gates that control a particular $\varepsilon$, there are sufficient tunable gates in typical experimental devices to allow us  assume independent control over each  parameter. Varying detuning is preferred over tunnel coupling as it is generally easier to achieve in experiments. Our schemes are viable to quantum dots in Si~\cite{Zwanenburg:2013p961} because Si has small spin-orbit interaction~\cite{Tahan:2002p035314}, low proportion of spinful nuclei ($5\%$ of spin-1/2 $^{29}$Si), and can be further isotopically purified. Therefore, our schemes are feasible and realistic for current experiments.

\section{Theoretical Model}\label{sec:theory}

\subsection{Hamiltonian}
We consider a chain of 3 quantum dots with nearest neighbour couplings, described by a Hubbard model~\cite{Yang:2011p161301, DasSarma:2011p235314}, 
\begin{equation}
	H = H_{\mu} +  H_t + H_U,
	\label{eqn:H2}
\end{equation}
where $H_\mu$ is a term that depends on the electrochemical potential of the quantum dots and its detuning energy, $H_t$ is the term describing the nearest neighbour inter-dot hopping, and $H_U$ describes both intra-dot and inter-dot Coulomb interactions. (See Appendix~\ref{appendix:hubbard} for details.)

Eq.~\ref{eqn:H2} assumes  each quantum dot is either empty, singly or doubly occupied with both electrons in the ground orbital forming a spin singlet. We assume that the single particle excited states are well separated from the ground state. These excited states may be orbital or valley in nature~\cite{Zwanenburg:2013p961}. The lowest two valley states can be engineered to produce a gap of several meV~\cite{Rahman:2011p195323}, up to 9~meV~\cite{Zhang:2013p2396}. For orbital states, the gap can be as large as 8~meV based on experimentally measured orbital spectra in Si/SiGe quantum dots~\cite{PhysRevB.86.115319}. Experiments are carried out at 100~mK temperatures, so thermal excitations may be neglected. Also, the effect of higher orbitals causes only a small re-normalisation of the Hubbard parameters~\cite{Wang:2011p115301}. For these reasons, we retain only ground orbitals in our model.

For the ST qubit, an inter-dot magnetic field difference $\Delta B$ arises from nuclear spins~\cite{Maune:2012p344} or micro-magnets~\cite{Wu:2014p11938, PioroLadriere:2008p776}, and offers control over two independent rotation axes on the qubit Bloch sphere. While there is no direct control over $\Delta B$, electrical control over detuning allows the qubit to be pulsed quickly into regimes where either exchange or magnetic coupling energies dominate. In this work, we consider the regime where the exchange energy dominates. The value of exchange we adopt is several orders of magnitude larger than typical experimental values of the magnetic energy term $g\mu_\text{B}\Delta B$ in Si: $\sim3$~neV in natural Si~\cite{Assali:2011p165301} or $\sim60$~neV with micro-magnets~\cite{Wu:2014p11938}. Here, $g$ is the $g$-factor of the host semiconductor; $\mu_\text{B}$ is the Bohr magneton. Also, typical in-plane magnetic fields affects the Zeeman energy of spin states outside the $S_z=0$ basis, thus we exclude  magnetic terms in Eq.~\ref{eqn:H2}.

In our numerical simulations, we use typical quantum dot parameters extracted from fits to a Hubbard Hamiltonian in Refs.~\onlinecite{DasSarma:2011p235314, Wang:2011p115301}, which are based on data from experiments in Si/SiGe quantum dots~\cite{Simmons:2009p3234}. We assume identical dots with identical, constant nearest-neighbour tunnel couplings, $t=0.12$ meV~\cite{DasSarma:2011p235314, Wang:2011p115301}. We take identical intra-dot and nearest neighbour inter-dot Coulomb energies of $U_i = U = 6.1$ meV~\cite{DasSarma:2011p235314} and $U_{ij} = K$ respectively.

% Schemes for State Transfer
\subsection{Fidelity of State Transfer}

We propose two schemes for state transfer. The first is a pulse-gated scheme where dc pulses control the detuning. The second is an adiabatic scheme where detuning is changed linearly through energy anti-crossings. Our aim is to coherently transfer an arbitrary superposition of the $S$ and $T_0$ states of the ST qubit from the leftmost dots, 1 and 2, to the rightmost dots, 2 and 3. Because the Hamiltonian is spin conserving, it is block diagonal in spin space. Although the singlet and triplet states are uncoupled, the schemes do not require knowledge of the initial admixture of singlet and triplet states. This arbitrary superposition of states leads to the problem of state transfer becoming non-trivial because of the non-identical inter-dot coupling of the different spin and charge states, errors from phase accumulation, and leakage into states of undesired charge occupation.

We start with a general arbitrary initial state in the first two quantum dots, $\ket{\psi_0} = \cos \theta \ket{S}_{1,2} + e^{i \phi} \sin \theta \ket{T_0}_{1,2}$, where $\Ket{S/T_0}_{i,j} = (\Ket{\uparrow_i \downarrow_j}\mp \Ket{\downarrow_i \uparrow_j})/\sqrt{2}$. The mixing angle $\theta$ determines the admixture of the singlet and triplet states, and $\phi$ is the initial phase difference. The aim is to obtain the target state $\ket{\psi_\text{tgt}} = \cos{\theta} \ket{S}_{2,3} +  e^{i \phi} \sin{\theta} \ket{T_0}_{2,3}$, for the triple dot chain. 

The key figure of merit is the fidelity of state transfer~\cite{Loss:1998p120,flentje_2017,fujita_2017,Christandl:2004p187902,Vermersch_2017,Shulman:2012p202,PhysRevB.91.205434,Kawakami:2014p666,PhysRevLett.103.110501} as functions of the initial mixing angle and phase. While there are other measures of state transfer quality, fidelity is intuitively simple;
it is a measure of how close we are to achieving the target state~\cite{Nielsen:2000}. We report the best and worst  fidelities (see Tables~\ref{table:pulse}, \ref{table:adiabatic} and Appendix~\ref{appendix:fidcalc}) as well as the fidelity averaged over the entire range of mixing angles (see Appendix~\ref{appendix:avgfid}). Calculations over the entire range of initial mixing angles and phases indicate that fidelity contains no dependence on the initial phase (see Appendix~\ref{appendix:phi}). We therefore present results for one initial phase angle $\phi=0$, in the main text; identical results apply for other values of $\phi$.

\section{State Transfer Simulations}

\subsection{Master Equation}
In all our numerical simulations, we use 9 basis states, comprising 3 different charge states within the unpolarised triplet spin space 
and 6 singly- and doubly-occupied charge states within the singlet spin space (see Appendix~\ref{appendix:hubbard}). 

We simulate the state transfer outlined in Section~\ref{sec:theory} by solving the $9\times 9$ density matrix for our three dot, two electron system using a Markovian master equation~\cite{Barrett:2002p125318,PhysRevB.91.205434} 
\begin{equation}
	\frac{\partial \rho(\tau)}{\partial \tau} = -\frac{i}{\hbar} [H(\tau), \rho(\tau)] -  D[\rho(\tau)],
	\label{eqn:master}
\end{equation}
where $\tau$ is time, the first term on the right hand side describes the coherent evolution and the second term describes dephasing effects. The latter is explained in Section~\ref{sec:dephase} next, and Appendix~\ref{appendix:dephasingcalc} details the calculations.

% Dephasing
\subsection{Dephasing}\label{sec:dephase}

We now examine the effects of orbital dephasing in the solid state environment. We assume weak coupling between the system and a bosonic environment which may consist of phonons or charge degrees of freedom (``charge noise''). We assume that charge noise induces uncorrelated variations in the detuning parameter and result in fluctuations in the energy differences in the system, leading to dephasing effects~\cite{Barrett:2002p125318,PhysRevB.91.205434}. Although it is possible that charge noise affects the tunnel coupling parameter, we do not specifically include this as it is not thought to be a dominant noise source~\cite{PhysRevLett.110.146804}.

Charge noise can be characterised by two types of noise: double occupation dephasing noise~\cite{Barrett:2002p125318} and single occupation dephasing noise~\cite{PhysRevB.76.035315}. Double-occupation dephasing is formulated in Eq.~\ref{eqn:master} as
$D_D = \sum_i \frac{\Gamma}{2} \comm{n_{i\uparrow}+n_{i\downarrow}}{\comm{n_{i\uparrow}+n_{i\downarrow}}{\rho}}$,
where $n_{i\sigma}$ is the electron number operator with spin $\sigma$, $\rho$ is the density matrix we want to solve, and $\Gamma$ is the dephasing rate, which we take to be 1 GHz~\cite{PhysRevB.91.205434}. This term is attributed to virtual transitions to the doubly occupied states 
 %($\ket{S}_{1,1},\ket{S}_{2,2},\ket{S}_{3,3}$) 
during the state transfer.  
Single occupation dephasing is given by 
$D_S = \sum_{i,j}{\gamma_{ij}\rho_{ij}(\tau) |i\rangle\langle j|}$, where $i \neq j$,
which describes dephasing between singly-occupied states at rates $\gamma_{ij}$ that depend on the energy splitting of the states $E_{ij}$ with respect to detuning of the leftmost and rightmost dots (1 and 3), i.e.
$\gamma_{ij} = \Gamma\sqrt{\left(\partial E_{ij}/\partial \varepsilon_1\right)^2+\left(\partial E_{ij}/\partial \varepsilon_3\right)^2}$, where $E_{ij} \equiv E_i-E_j$. The second term on the right hand side of Eq.~\ref{eqn:master} is defined as the sum of both dephasing terms, $D[\rho(\tau)] \equiv D_D + D_S$.

% pulse-gated State Transfer
\section{Pulse-Gated State Transfer}

\begin{figure}
    \includegraphics[width=3.4in]{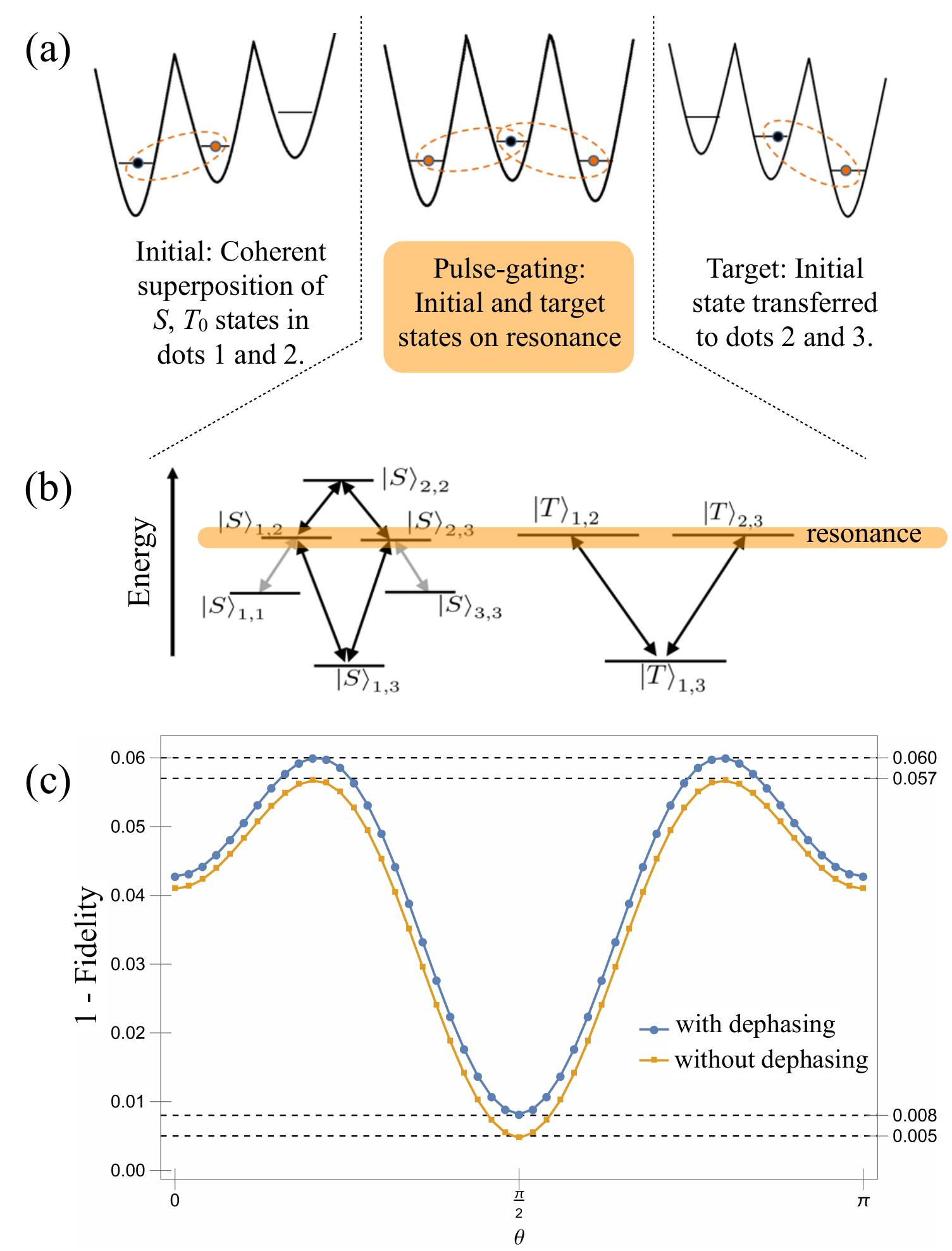}
	\caption{
	Pulse-Gated State Transfer.
	(a) Schematic of the detunings before, during and after the pulse. Electrons are represented by filled circles; black horizontal lines represents the ground orbitals. In the left and right regimes, the two lowest lying states are the $S$ and $T_0$ states of the two leftmost and rightmost dots respectively as represented by dashed ellipses. In the middle regime, the left and right dots are on resonance, allowing state transfer between the $\Ket{S,T_0}_{1,2}$ and $\Ket{S,T_0}_{2,3}$ states (dashed ellipses).
	(b) Resonance condition during pulse-gating. Coupling via the intermediate states are shown in black arrows, while coupling to leakage states of undesired charge occupation are shown in grey arrows. These arrows showing tunnel couplings are sketched against a vertical energy axis (not to scale).
	(c) Plot of infidelity $(1- \text{fidelity}~\mathcal{F})$ with mixing angle $\theta$ with dephasing (red circles) and without (yellow squares),  where $U=2K$, $K=3.05$~meV and $t=0.12$~meV. The upper and lower bounds for the values of infidelity are labelled in the plot as well. We use $\Gamma = 1$~GHz in the simulations.
}
	\label{fig:PulseGated}
\end{figure}

In the pulse-gated state transfer scheme, dc pulses are applied to control detuning $\varepsilon_i$. The pulse sequence moves the dots through several regimes (Fig. \ref{fig:PulseGated}(a)): first, the right dot is far detuned from the other dots ($|\varepsilon_3| \gg \varepsilon_1,\varepsilon_2$), next the left and right dots are moved such that they are on resonance, but detuned from the middle dot ($ \varepsilon_2 > \varepsilon \equiv \varepsilon_1 = \varepsilon_3$) and finally a regime where the left dot is far detuned from the other dots ($|\varepsilon_1| \gg \varepsilon_3,\varepsilon_2$).

%%%%%%%%%%%%%%%%%%%%%%%%%%
%%%%%%%%%%%%%%%%%%%%

We bring the $\ket{S/T_0}_{1,2}$ and $\ket{S/T_0}_{2,3}$ states into resonance with each other via control over the detuning energies of dots 1 and 3 as shown in the schematics of Fig.~\ref{fig:PulseGated}(b) and the middle panel of Fig.~\ref{fig:PulseGated}(a). The initial and target states for the singlet are coupled via intermediate states $\ket{S}_{1,3}$ and $\ket{S}_{2,2}$, while those for the triplet are coupled via $\ket{T_0}_{1,3}$, as shown in black arrows in Fig.~\ref{fig:PulseGated}(b). The singlet states also couple to leakage states of undesired charge occupation $\Ket{S}_{1,1}$ and $\Ket{S}_{3,3}$, shown in grey arrows. 

The Schrieffer-Wolff transformation~\cite{Schrieffer:1966p491, Gros:1987p381, MacDonald:1988p9753} allows us to gain insight into the effective couplings between the initial and target states. We can obtain the couplings to be $J_S = t^2 \left(\frac{-4}{U-K+\varepsilon}+\frac{2}{K+\varepsilon}\right)$, and $J_T = t^2 \left(\frac{2}{K+\varepsilon}\right)$ 
for the singlet and triplet states respectively. This suggests that for a pure singlet or triplet state, the target state may be reached with the highest fidelity after a gate time of $\tau_\text{gate} = \hbar \pi/J_{S/T}$ on resonance. However, this gate time differs for the singlet and triplet states if $J_S$ and $J_T$ are different. Therefore, for an arbitrary initial state, we require the gating times to be equal,  $\abs{J_S}=\abs{J_T}\equiv J$, in order to achieve maximum fidelity with  simultaneous transfer of both spin states. This can be satisfied if the ratio of the intra-dot to inter-dot Coulomb repulsion is given by $U/K=2$. This condition corresponds to a realistic constraint on the ratio of inter-dot distance to dot size, and is detailed in Appendix \ref{appendix:feasible}. We emphasise here again that the Schrieffer-Wolff Hamiltonian was useful for gaining insight into the gating time, and was not used in the simulations since it is effectively an approximation.

We also note that the lowest-lying excited states must be well separated from those involved in the pulse-gating scheme to avoid undesired resonances. For singlets, this means that the energy difference between the doubly-occupied (2,2) singlet and singly-occupied (1,2) and (2,3) excited states must be $\Delta E_{\text{S}} = K - U - \varepsilon + E_{ex}$, where $E_{ex}$ is the single particle excited energy in each dot which we take to be 8 meV \cite{PhysRevB.86.115319}. For triplet states, the excited (1,3) state must be much higher than the energies of the (1,2) and (2,3) states, with a gap of $\Delta E_\text{T} = \varepsilon - K + E_{ex}$.

After solving Eq.~\ref{eqn:master} with the pulses, we obtain the final state which we then use for the calculation of fidelity, which are shown in Table \ref{table:pulse}.
In Fig.~\ref{fig:PulseGated}(c), we plot the infidelity ($1-\mathcal{F}$) against various mixing angles $\theta$ of the initial state, with $\phi=0$, for both cases -- without  and with dephasing. The best fidelity is achieved when the state vector is a pure triplet state, and the inclusion of the dephasing rate $\Gamma = 1$~GHz, reduces fidelity across  $\theta$. 

\begin{center}
\begin{table}
    \begin{tabular}{|>{\centering\arraybackslash} m{0.25\columnwidth}|>{\centering\arraybackslash} m{0.225\columnwidth}|>{\centering\arraybackslash} m{0.225\columnwidth}|>{\centering\arraybackslash} m{0.225\columnwidth}|}
    \hline
    \textbf{Pulse-Gated} & Worst  & Best  & Average  \\
    \textbf{Scheme} & Fidelity & Fidelity & Fidelity \\
    \hline
    Without Dephasing & 94.3\% & 99.5\% & 96.2\% \\
    \hline
    With Dephasing & 94.0\% & 99.2\% & 95.8\% \\
    \hline
    \end{tabular}
\caption{Fidelities for the pulse-gated scheme. $\Gamma=1$~GHz was used in calculations with dephasing.}
\label{table:pulse}
\end{table}
\begin{table}
    \begin{tabular}{|>{\centering\arraybackslash} m{0.225\columnwidth}|>{\centering\arraybackslash} m{0.225\columnwidth}|>{\centering\arraybackslash} m{0.225\columnwidth}|>{\centering\arraybackslash} m{0.225\columnwidth}|}
    \hline
    \textbf{Adiabatic} & Worst  & Best  & Average  \\
    \textbf{Scheme} & Fidelity & Fidelity & Fidelity \\
    \hline
    Without Dephasing & 75.4\% & 99.0\% & 84.8\% \\
    \hline
    With Dephasing & 64.9\% & 93.6\% & 77.6\% \\
    \hline
    \end{tabular}
\caption{Fidelities for the adiabatic scheme. $\Gamma=1$~GHz was used in calculations with dephasing.}
\label{table:adiabatic}
\end{table}
\end{center}

%%%%%%%%%%%%%%%%%%%%%%%%%%
\begin{figure}
	\includegraphics[width=3.3in]{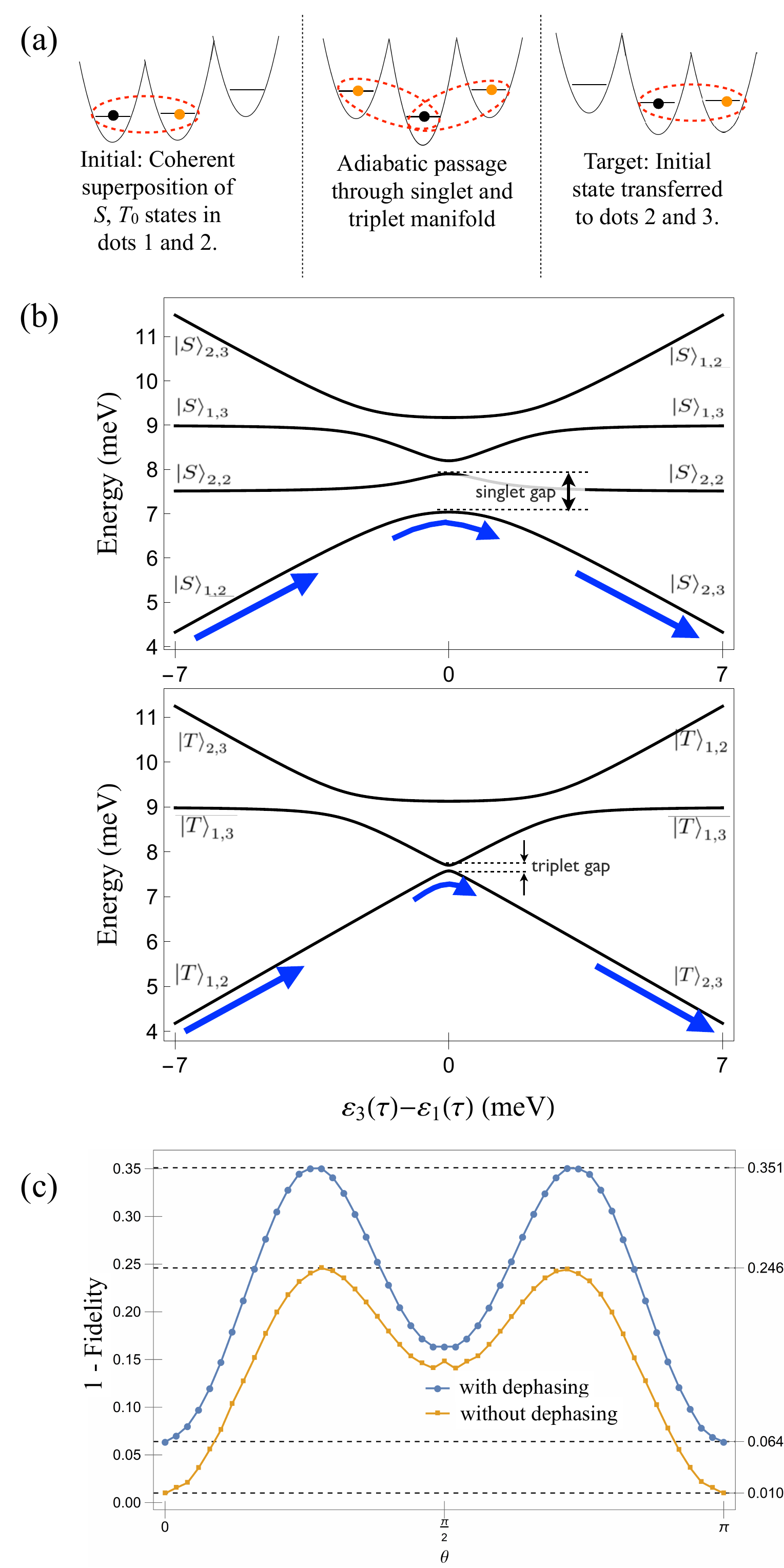}
	\caption{
	Adiabatic State Transfer.
	(a), (b) At far-detuned initial and target regimes, charge occupation of the dots are good eigenstates, and the system starts in an arbitrary superposition of the ground $\Ket{S,T_0}_{1,2}$ states. As the detuning is swept along the direction of the red arrows, the system evolves adiabatically to $\Ket{S, T_0}_{2,3}$, with singlet and triplet states picking up different dynamical phase contributions. Here, the energies of the singlet and triplet states are plotted as a function of $\varepsilon_3(\tau)-\varepsilon_1(\tau)$. The detuning is applied on the first and third dots as a simple linear ramp, while the detuning for the second dot is kept constant. (See Appendix~\ref{appendix:detuning} for details.)
	(c) The plot of infidelity $(1- \text{fidelity}~\mathcal{F})$ with mixing angle $\theta$ with dephasing (red circles) and without (yellow squares) ,  where $U=6.1$ meV, $K=2.5$~meV and $t=0.12$~meV. The upper and lower bounds for the values of infidelity are also labelled in the plot. Similarly, the dephasing rate, $\Gamma$, is given by 1 GHz.
	}
	\label{fig:Adiabatic}
\end{figure}
%%%%%%%%%%%%%%%%%%%%%%%%%%

% ADIABATIC STATE TRANSFER
\section{Adiabatic State Transfer}

In the adiabatic scheme, the initial state adiabatically evolves to the desired target state by tuning the detuning energies at a rate that the system remains in the ground state through the anti-crossings, as shown in the schematics of Fig.~\ref{fig:Adiabatic}(a, b). This scheme provides an alternative to pulse-gating in the case where engineering the ratio $U/K=2$ is challenging. We therefore consider a different inter-dot Coulomb interaction strength of $K=2.5$~meV~\cite{DasSarma:2011p235314},  keeping other parameters the same.

The avoided crossings occur around $\varepsilon_3(\tau)=\varepsilon_1(\tau)$, as shown in Fig.~\ref{fig:Adiabatic}(b). We observe that avoided crossings for singlets arises from the tunnel couplings between $\Ket{S}_{1,2}$ and $\Ket{S}_{2,3}$ with $\Ket{S}_{2,2}$, while the avoided crossing for triplets (Fig.~\ref{fig:Adiabatic}(b)) arises from the tunnel couplings  between $\Ket{T_0}_{1,2}$ and $\Ket{T_0}_{2,3}$ with $\Ket{T_0}_{1,3}$. Therefore, the energy gap for singlets and triplets are respectively, of the order $\varepsilon_{\mathrm{gap,S}}\sim\mathcal{O}(t)$, and $\varepsilon_{\mathrm{gap,T}}\sim\mathcal{O}(t^2/(\varepsilon+K))$, with $\varepsilon_{\mathrm{gap,T}}<\varepsilon_{\mathrm{gap,S}}$. The smaller of the two gaps, $\varepsilon_{\mathrm{gap,T}}$ sets an lower bound on the duration for adiabatic passage. 

In this scheme, detuning $\varepsilon_1$ and $\varepsilon_3$ are evolved at a constant rate over a duration $R$ such that $\varepsilon_1 = \varepsilon_3$ at time $\tau = R/2$ while $\epsilon_2$ remain detuned. This is shown schematically in Fig.~\ref{fig:Adiabatic}. (See Appendix~\ref{appendix:detuning}.) The strategy we employ is as follows: first, we estimate $R$ based on dephasing estimates, then using this $R$, we determine the optimal state transfer time from average fidelity calculations (see Fig.~\ref{fig:fidtime} in Appendix.~\ref{appendix:avgfid}). The latter step is needed because precise dephasing rates in real experiments are typically unknown.

To estimate the duration $R$, we make use of the adiabaticity condition, $R \gg \hbar (-\ln P_D) (\varepsilon-1)/ 2\pi t^2 $, based on a simple Landau-Zener model~\cite{LZ-ref}. Here, $P_{D}$ is the threshold probability tolerable for the state transfer, which we set to be negligibly small. This sets the duration, which we calculate to be $R = 9.9$~ns. In the absence of noise, an infinitely long duration would yield arbitrarily high fidelities; our choice of finite $R$ is of the same order of magnitude or less than dephasing times in isotopically natural silicon dots, which range from $\sim10~\mathrm{ns}$ to 900 ns~\cite{Maune:2012p344, Shi:2013p3020,Kawakami:2014p666}. We argue that this value of $R$ is reasonable since we expect realistic adiabatic durations to be bounded by dephasing times. (In isotopically purified silicon dots, the dephasing time is much longer, at 2.31~$\mu$s~\cite{Eng:2015pe1500214}.)

We find that the best fidelities occur after a transfer time of about 6.9 ns, which is approximately of the correct order that we anticipated above\cite{PhysRevLett.103.110501}.  The fidelities obtained at that instant are shown in Table \ref{table:adiabatic}. Fig. \ref{fig:Adiabatic}(c) contains the plots of the infidelity $(1-\mathcal{F})$ across $\theta$ with $\phi=0$ for both cases with and without dephasing.

\section{Discussion}

Comparing the fidelities in Tables \ref{table:pulse} and \ref{table:adiabatic}, we see that the highest fidelity achieved for both schemes with or without dephasing are comparable. However, the lowest fidelity is significantly better for pulse-gating than the adiabatic scheme. This is not surprising because in any adiabatic scheme, there is a natural trade-off between speed and adiabaticity --  too slow and dephasing reduces quality; too fast and diabatic transitions reduce fidelity. 

In contrast, pulsed gate fidelities are not reduced as significantly with dephasing, as long as the rise time is sufficiently short to satisfy diabaticity (see Appendix~\ref{appendix:diabaticity}), and when the gate time is much faster than dephasing time. The latter condition is favourable when tunnel coupling is large, as it is the case in our calculations. On the other hand,  a large tunnel coupling sets an upper bound on the pulse rise time; a short rise time may be challenging to implement, depending on the signal generator bandwidth. This can be overcome by a suitable reduction of the tunnel coupling: in experiments, there are typically sufficient gates that tunnel coupling is tunable over a wide range~\cite{PhysRevB.72.081310, Thalakulam:2010p183104, Simmons:2009p3234}.

One of the challenges in determining the best fidelity in the adiabatic scheme comes from the non-trivial oscillations in fidelity that occur right after adiabatic passage through the energy gaps (see Fig.~\ref{fig:fidtime} of Appendix~\ref{appendix:avgfid}). A complete analysis of the dynamical evolution of fidelity is out of the scope of this paper, however. We surmise that different phase accumulation for triplet and singlet states and the coupling of the target state to the multiple levels in the system produce these oscillations.  The consequence is that the adiabatic scheme will require careful experimental calibration with known initial states for the optimization of fidelity.

% SUMMARY
\section{Conclusions}

In summary, we  presented two schemes for ST qubit state transfer in a chain of 3 quantum dots. That these schemes are feasible and accessible to current experiments is central to addressing the practical implementation of the transfer of quantum information, so as to advance the scalability of ST qubits. The scalability of ST qubits will be the next step in our implementation of the coherent transfer of the coherent state. The key question will be how much the fidelity of state transfer will be affected if the two schemes we discussed here are to be repeated over a chain of arbitrarily many quantum dots, instead of just 3 in our scenario.

In our scheme of three dots, the pulse-gated scheme gives fidelities between $94.3\% <\mathcal{F}< 99.5\%$ across $\theta$ within a short transfer time of 0.076~ns, by requiring $U/K=2$. Adiabatic state transfer achieves fidelities between $75.4\% <\mathcal{F}< 99.0\%$ at a longer time (6 ns), but without the condition on the Coulomb energies ratio. Taking into account dephasing effects, the values of fidelity are now between $94.0\% <\mathcal{F}< 99.2\%$ for the pulse-gated scheme and between $64.9\% <\mathcal{F}< 93.6\%$ for the adiabatic scheme.

% ACKNOWLEDGEMENT

\begin{acknowledgments}
We thank David P. DiVincenzo for helpful discussions. MK Feng gratefully acknowledges support by Singapore Ministry of Education Academic Research Fund (Grant No. RG117/16). We would also like to thank the anonymous referees for their valuable and insightful comments.
\end{acknowledgments}

\appendix
\renewcommand\thefigure{\thesection.\arabic{figure}}  

% Hubbard Model
\section{Hamiltonian}
\label{appendix:hubbard}

The Hamiltonian in Eq.~\ref{eqn:H2} of the main text contains the following terms,
\begin{align}
	H_\mu&=- \sum_{i,\sigma} (\mu_i + \varepsilon_i\left( \tau \right)) n_{i, \sigma}, \\
	H_t&= -\sum_{<i,j>,\sigma} t c^\dag_{i\sigma} c_{j \sigma},  \\
	H_U&= \sum_{i} U_i n_{i,\uparrow}n_{i,\downarrow} +\frac{1}{2}\sum_{<i,j>} U_{ij}( n_{i,\uparrow}n_{j,\downarrow}+ n_{j,\downarrow}n_{i,\uparrow}).
\end{align}
where $\mu_i$ is the electrochemical potential of the $i$-th dot and $\varepsilon_i(\tau)$ is the experimentally controlled detuning energy of the $i$-th dot with time $\tau$, $n_{i,\sigma}=c_{i\sigma}^\dagger c_{j\sigma}$ is the electron number operator on the $i$-th dot with spin $\sigma$, $t$ is tunnel coupling, $U_i$ describes the intra-dot Coulomb energy and $U_{ij}$ describes the inter-dot direct Coulomb energy. In our model, we take the inter-dot direct Coulomb energy to be equal between each dot, $U_{12}=U_{23}$.
%The Hamiltonian can be categorised into Coulomb interaction consisting of the inter-dot and intra-dot Coulomb interaction with the tunnelling, exchange interaction and the detunings treated as perturbation. 

The basis singlet states are
\begin{align}
	\ket{1} &\equiv \Ket{S}_{3,3} = \Ket{\uparrow_3 \downarrow_3},  \\
   \ket{2} &\equiv \Ket{S}_{2,3} = \frac{1}{\sqrt{2}}\left(\Ket{\uparrow_2 \downarrow_3}-\Ket{\downarrow_2 \uparrow_3}\right), \\
	\ket{3} &\equiv \Ket{S}_{1,3} = \frac{1}{\sqrt{2}}\left(\Ket{\uparrow_1 \downarrow_3}-\Ket{\downarrow_1 \uparrow_3},\right)\\
	\ket{4} &\equiv \Ket{S}_{2,2} =\Ket{\uparrow_2 \downarrow_2},\\
	\ket{5} &\equiv \Ket{S}_{1,2} = \frac{1}{\sqrt{2}}\left(\Ket{\uparrow_1 \downarrow_2}-\Ket{\downarrow_1 \uparrow_2}\right), \\
	\ket{6} &\equiv \Ket{S}_{1,1} =\Ket{\uparrow_1 \downarrow_1},
\end{align}
while the basis states for triplets are
\begin{align}
	\ket{7} &\equiv \Ket{T_0}_{1,2} = \frac{1}{\sqrt{2}}\left(\Ket{\uparrow_1 \downarrow_2}+\Ket{\downarrow_1 \uparrow_2}\right), \\
	\ket{8} &\equiv \Ket{T_0}_{1,3} = \frac{1}{\sqrt{2}}\left(\Ket{\uparrow_1 \downarrow_3}+\Ket{\downarrow_1 \uparrow_3}\right), \\
	\ket{9} &\equiv \Ket{T_0}_{2,3} = \frac{1}{\sqrt{2}}\left(\Ket{\uparrow_2 \downarrow_3}+\Ket{\downarrow_2 \uparrow_3}\right).
\end{align}

The Hamiltonian is block diagonal in spin space and the singlet and triplet blocks are given by
\begin{widetext}
	\begin{align}
		\hat{H}_{S}&=
		\begin{pmatrix}
			U-2(\varepsilon_3+\mu)	&	-\sqrt{2}t	&0	&0	&0	&0\\
			-\sqrt{2}t	&	U_{12}-\varepsilon_2-\varepsilon_3-2\mu	&-t	&-\sqrt{2}t &0	&0 \\
			0	&	-t	&-(\varepsilon_1+\varepsilon_3+2\mu) 	&0	&-t 	&0\\
			0	&	-\sqrt{2}t 	&0	&U-2(\varepsilon_2+\mu) 	&-\sqrt{2}t 	&0\\
			0	&0	&-t	&-\sqrt{2}t 	&U_{12}-\varepsilon_1-\varepsilon_2-2\mu 	&-\sqrt{2}t \\
			0	&0	&0	&0	&-\sqrt{2}t 	&U-2(\varepsilon_1+\mu)\\
		\end{pmatrix}, \\
		\hat{H}_{T}&=
		\begin{pmatrix}
			U_{12}-\varepsilon_1-\varepsilon_2-2\mu 	&-t 	&0\\
			-t	&-(\varepsilon_1+\varepsilon_3+2\mu) 	&-t\\
			0& -t 	&U_{12}-\varepsilon_2-\varepsilon_3-2\mu\\
		\end{pmatrix},
	\end{align}
\end{widetext}
where $\hat{H}_{S}$ is the Hamiltonian for singlet states and $\hat{H}_{T}$ is the Hamiltonian for triplet states. Together, they form a block diagonal Hamiltonian $\hat{H}$ since the singlet and triplet states do not mixed due to the absence of spin-orbit coupling and magnetic field.

%Detuning as a function of time

\section{Detuning pulses}
\label{appendix:detuning}

The detuning pulses used in our simulations are given here. For pulse-gating simulations, we used dc pulses with smoothly rising and falling steps, in a total time ranging from  $\tau_\text{start}-\tau_\text{rise}$ to $\tau_\text{end}+\tau_\text{rise}$, given by
\begin{widetext}
\begin{align}
	\varepsilon_1 (\tau) &= \begin{cases}
	(\varepsilon+\varepsilon_d) & \tau < \tau_\text{start}-\tau_\text{rise} \\
	-\varepsilon_d \cos[\omega(\tau-\tau_\text{start})] + (\varepsilon+\varepsilon_d) & \tau_\text{start}-\tau_\text{rise} < \tau < \tau_\text{start} \\
	\varepsilon & \tau_\text{start} < \tau < \tau_\text{end} \\
	\varepsilon_d \cos[\omega(\tau-\tau_\text{end})] + (\varepsilon-\varepsilon_d) & \tau_\text{end} < \tau < \tau_\text{end}+\tau_\text{rise} \\
	(\varepsilon-\varepsilon_d) & \tau > \tau_\text{end}+\tau_\text{rise}
	\end{cases},  \\
	\varepsilon_2 (\tau) &= \varepsilon, \\
	\varepsilon_3 (\tau) &= \begin{cases}
	(\varepsilon-\varepsilon_d) & \tau < \tau_\text{start}-\tau_\text{rise} \\
	\varepsilon_d \cos[\omega(\tau-\tau_\text{start})] + (\varepsilon-\varepsilon_d) & \tau_\text{start}-\tau_\text{rise} < \tau < \tau_\text{start} \\
	\varepsilon & \tau_\text{start} < \tau < \tau_\text{end} \\
	 -\varepsilon_d \cos[\omega(\tau-\tau_\text{end})] + (\varepsilon+\varepsilon_d) & \tau_\text{end} < \tau < \tau_\text{end}+\tau_\text{rise} \\
	 (\varepsilon+\varepsilon_d)& \tau > \tau_\text{end}+\tau_\text{rise}
	\end{cases},
\end{align}
where $\varepsilon=-2$ meV and $\varepsilon_d=3$ meV.
\end{widetext}
For the adiabatic scheme,  detuning pulses used in simulations are
\begin{align}
	\varepsilon_1 (\tau) &= \begin{cases}
	\varepsilon_\text{high} & \tau < 0 \\
	\frac{(\varepsilon_\text{low}-\varepsilon_\text{high})\tau}{R}+\varepsilon_\text{high} & 0 < \tau < R \\
	\varepsilon_\text{low} & \tau > R
	\end{cases},  \\
	\varepsilon_2 (\tau) &= \varepsilon_\text{high}, \\
	\varepsilon_3 (\tau) &= \begin{cases}
	\varepsilon_\text{low} & \tau < 0 \\
	\frac{(\varepsilon_\text{high}-\varepsilon_\text{low})\tau}{R}+\varepsilon_\text{low} & 0 < \tau < R \\
	\varepsilon_\text{high} & \tau > R
	\end{cases},
\end{align}
where $\varepsilon_\text{low}=-8$ meV and $\varepsilon_\text{high}=-1$ meV.

For pulse-gating, the results we report are for a rise time $\tau_\text{rise} = \tau_\text{gate}/1000= 0.076$~ps to satisfy diabaticity, so that  it is effectively a square pulse. Although this imposes a requirement on experimental bandwidth capabilities, as explained in Appendix.~\ref{appendix:diabaticity}, it is not essential for obtaining good fidelities. In Fig.~\ref{fig:risetimecomparison}), we show that similar fidelity can be obtained for a much longer rise time, provided tunnel coupling is tuned to a smaller magnitude.

\section{Diabaticity of Pulse-Gated Rise Times}
\label{appendix:diabaticity}
\begin{figure}
	\includegraphics[width=3.4in]{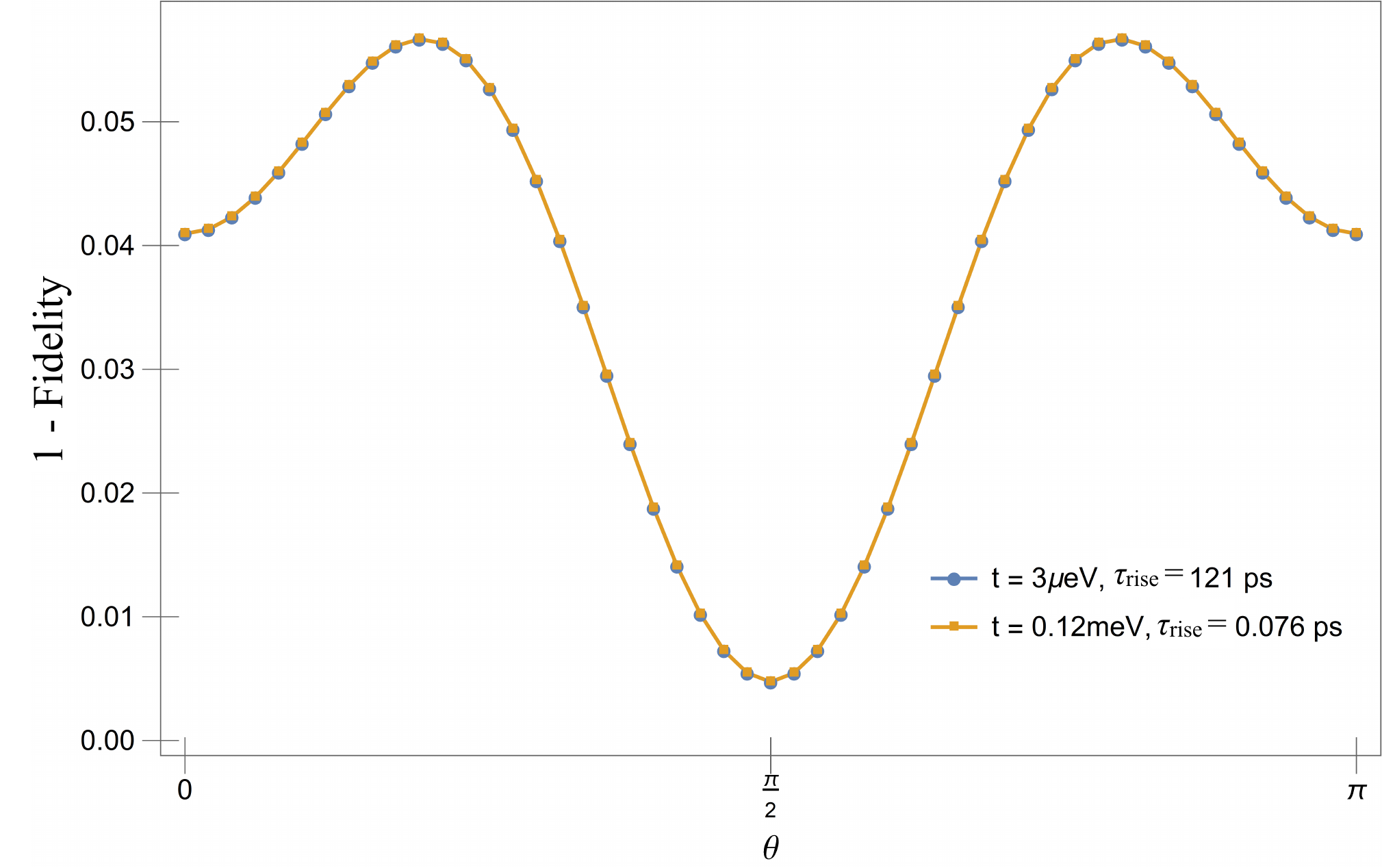}
	\caption{Comparison of infidelities (1 - fidelity $\mathcal{F}$) against mixing angle $\theta$ obtained for pulse-gating without dephasing, for two sets of pulse rise time and tunnel coupling. For tunnel coupling $t=0.12$~meV, we reported fidelity results  in Fig.~\ref{fig:Adiabatic} of the main text, using a short rise time $\tau_\text{rise} = 0.076$~ps, in order to simulate an effectively square pulse. A longer and more realistic rise time of $\tau_\text{rise} = 121$~ps can be achieved using a smaller tunnel coupling $t=3~\mu$eV to obtain identical fidelities. This demonstrates that rise time (or equivalently,  bandwidth of the signal generator) is not a limiting factor because tunnel coupling is tunable, as reported in Refs.~\onlinecite{PhysRevB.72.081310, Thalakulam:2010p183104, Simmons:2009p3234}.
    }
	\label{fig:risetimecomparison}
\end{figure}

For the pulse-gating scheme, detuning pulses may move the initial state through undesired anti-crossings in the energy landscape of the system during the rise and fall of the pulse. The rise time must be short enough that the evolution is effectively instantaneous. In the ideal limit, pulse-gating require instantaneous pulses. However, due to finite bandwidths of signal generators, rise times in real experiments are necessarily finite. The rate of change of detuning must satisfy $\frac{d\varepsilon}{d \tau} \gg t^2 /h$ in order to be effectively instantaneous. With $t=0.12$~meV used in our calculations, this leads to a rise time of less than 1~ps which is rather demanding for current experiments. If tunnel coupling is tuned by two orders of magnitude lower, e.g. $t= 3~\mu$eV, a more achievable rise time of 121~ps is needed, for identical fidelities. This is shown in Fig.~\ref{fig:risetimecomparison}. This demonstrates that rise time (or equivalently,  bandwidth of the signal generator) is not a limiting factor because of the tunability of tunnel coupling~\cite{PhysRevB.72.081310, Thalakulam:2010p183104, Simmons:2009p3234}.

%%%%%%%%%%%%%%%%%%%%%%%%%%
%%%%%%%%%%%%%%%%%%%%

% Fidelity Calculation
\section{Fidelity Calculation}
\label{appendix:fidcalc}

To obtain the solutions making use of the density matrix formalism, we first recast the definition of fidelity in density matrix form \cite{Nielsen:2000}:
\begin{equation*}
\mathcal{F}(\rho,\sigma) \equiv \tr\sqrt{\rho^{1/2}\sigma\rho^{1/2}}.
\end{equation*}
We can simplify this definition if one of the density matrices describes our initial and target states which are pure states. The below form is obtained when we consider the fidelity between a pure state $\ket{\psi}$ and an arbitrary state, $\rho$:
\begin{equation*}
\mathcal{F}(\ket{\psi},\rho) = \tr\sqrt{\bra{\psi}\rho\ket{\psi}\ket{\psi}\bra{\psi}} = \sqrt{\bra{\psi}\rho\ket{\psi}}.
\end{equation*}

The initial density matrix elements are: $\rho_{55}(0)=\cos^2 \theta$, $\rho_{77}(0)=\sin^2 \theta$, $\rho_{57}(0)=\sin \theta \cos \theta e^{-i\phi}$, $\rho_{75}(0)=\sin \theta \cos\theta e^{i\phi}$, corresponding to the initial state,
$\ket{\psi_0} = \cos(\theta)\ket{S}_{1,2} + e^{i\phi}\sin(\theta)\ket{T_0}_{1,2}$.
The target density matrix elements are: $\sigma_{22}(\tau)=\cos^2 \theta$, $\sigma_{99}(\tau)=\sin^2 \theta$, $\sigma_{29}(\tau)=\sin \theta \cos \theta e^{-i\phi}$, $\sigma_{92}(\tau)=\sin \theta \cos \theta e^{i\phi}$ corresponds to the target state,
$\ket{\psi_{\text{tgt}}} = \cos(\theta)\ket{S}_{2,3} + \sin(\theta)e^{i\phi}\ket{T_0}_{2,3}$.
Therefore, fidelity can be written as
\begin{multline*}
\mathcal{F} = \left[\cos^2 (\theta)\rho_{22}(\tau) + \sin^2 (\theta)\rho_{99}(\tau) + \right. \\ \left. \cos(\theta)\sin(\theta)e^{i\phi}\rho_{29}(\tau) + \cos(\theta)\sin(\theta)e^{-i\phi}\rho_{92}(\tau)\right]^{1/2}
\end{multline*}

% Average Fidelity

\section{Average Fidelity}
\label{appendix:avgfid}

\begin{figure}
	\includegraphics[width=3.4in]{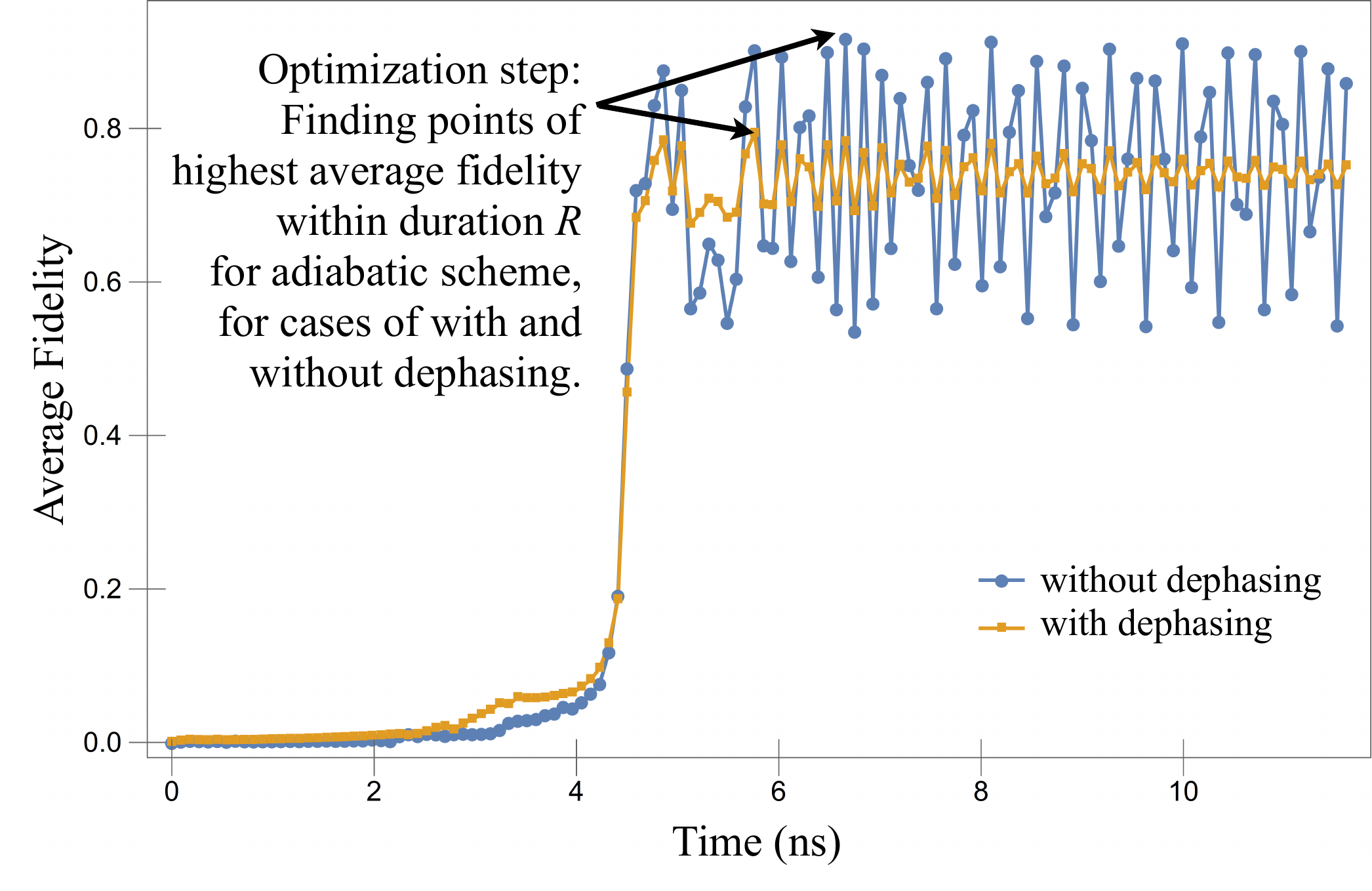}
	\caption{This figure depicts how the average values of fidelity in the adiabatic scheme evolves with time, for both the case without dephasing (red dots) and with dephasing (yellow squares). The average fidelity (Eq.~\ref{eqn:avgfid}) is constantly oscillating throughout the adiabatic transfer with a range of about 10\% for the case with dephasing and a range of about 20\% when without dephasing.
    }
	\label{fig:fidtime}
\end{figure}

The fidelity averaged over all mixing angles thus allows comparison of fidelities at different instants of time.
 Average fidelity is given by
\begin{equation}\label{eqn:avgfid}
\mathcal{F}_{\text{avg}}(\tau) = \frac{1}{2}\int^\pi_0 ~\mathcal{F(\theta, \tau)} \sin \theta ~\mathrm{d}\theta.
\end{equation}
Average fidelity was used to obtain the optimal adiabatic state transfer time, as illustrated in Fig.~\ref{fig:fidtime}.

%%%%%%%%%%%%%%%%%%%%%%%%%%
%%%%%%%%%%%%%%%%%%%%

% Show the Independence / Dependence of Phi

\section{Fidelity is Independent of Phase $\phi$}
\label{appendix:phi}

\begin{figure}[ht!]
	\includegraphics[width=3.2in]{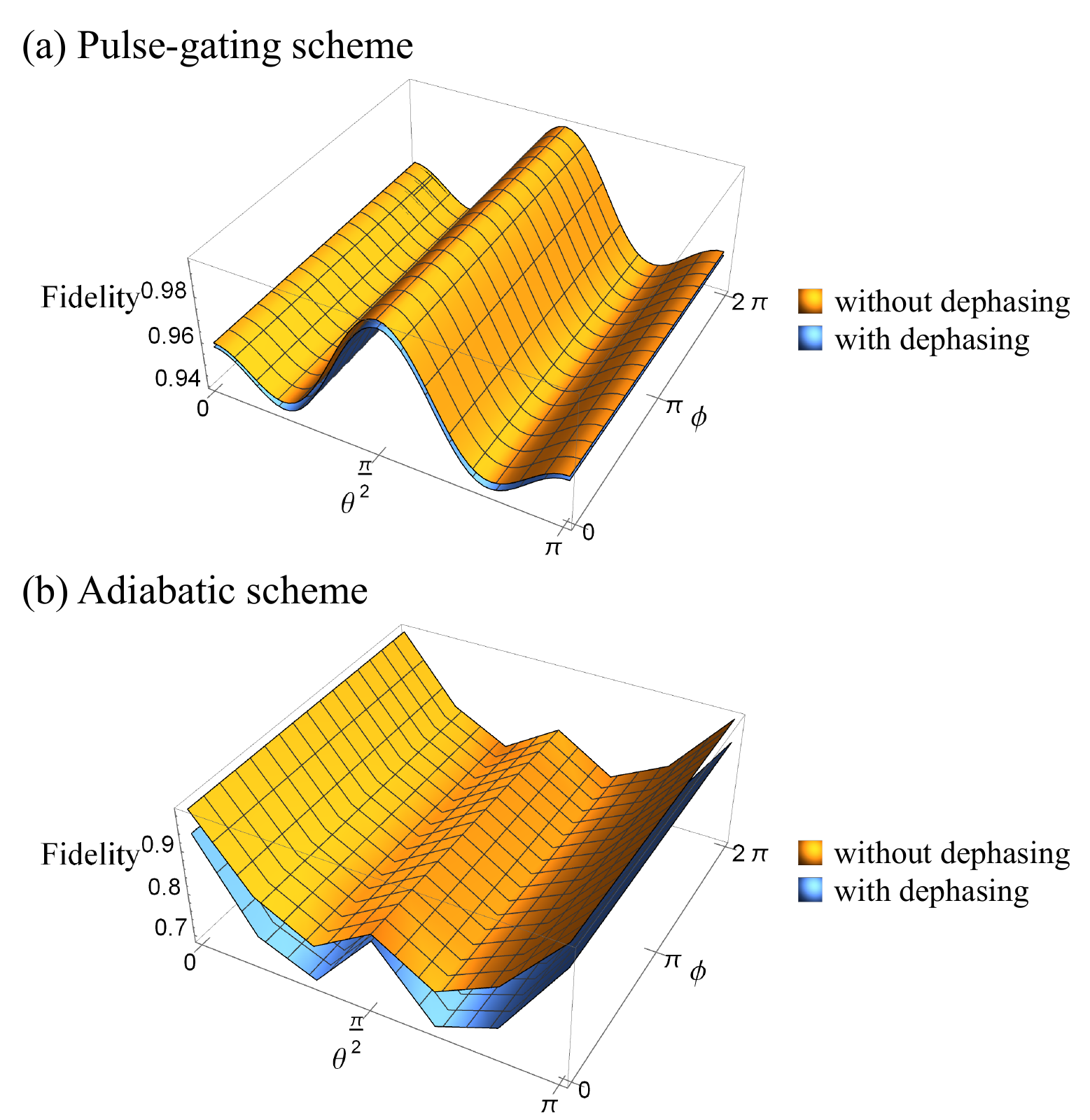}
	\caption{
	3D plot of fidelity with mixing angle $\theta$ and phase $\phi$.
	The maximum difference in fidelity across all initial phase angles is less than $10^{-7}$, indicating that fidelity is independent of initial phase angle $\phi$.
	}
	\label{fig:phi}
\end{figure}

In the main text, we reported results from simulations with initial phase angle $\phi = 0$. Here, we show numerical results, plotted in Fig.~\ref{fig:phi}, that indicate that fidelity is independent of $\phi$. We checked the fidelity calculated for the entire range of $0 \le \phi \le 2\pi$ for each mixing angle $\theta$ and found that the maximum difference in fidelity across initial phase angles is less than $10^{-7}$.

%%%%%%%%%%%%%%%%%%%%%%%%%%
%%%%%%%%%%%%%%%%%%%%%%%%%

% Dephasing Calculation

\section{Dephasing Terms}
\label{appendix:dephasingcalc}

As given in the main text, the double- and single-occupation dephasing terms in the master equation, Eq.~\ref{eqn:master} are given by
\begin{align}
 D_D &= \sum_i \frac{\Gamma}{2} \comm{n_{i\uparrow}+n_{i\downarrow}}{\comm{n_{i\uparrow}+n_{i\downarrow}}{\rho}},\\
 D_S &= \sum_{i,j}{\gamma_{ij}\rho_{ij}(\tau) |i\rangle\langle j|},
\end{align}
where $\Gamma = 1$~GHz is the double-occupation dephasing rate, $\gamma_{ij} = \Gamma\sqrt{\left(\partial E_{ij}/\partial \varepsilon_1\right)^2+\left(\partial E_{ij}/\partial \varepsilon_3\right)^2}$ is the single-occupation dephasing rate, and $E_{ij} \equiv E_i-E_j$ is the energy splitting. Note that we only consider the treatment of fluctuations of the energy splitting with respect  to $\varepsilon_1$ and $\varepsilon_3$ only, because $\varepsilon_2$ is held constant in both schemes.

Here, we write the dephasing matrices in the basis states explicitly. 
\begin{widetext}
\begin{equation*}
\begin{split}
D_D 
&= \begin{pmatrix}
0 & \Gamma\rho_{12}(\tau) & \Gamma\rho_{13}(\tau) & 4\Gamma\rho_{14}(\tau) & 3\Gamma\rho_{15}(\tau) & 4\Gamma\rho_{16}(\tau) & 3\Gamma\rho_{17}(\tau) & \Gamma\rho_{18}(\tau) & \Gamma\rho_{19}(\tau) \\
\Gamma\rho_{21}(\tau) & 0 & \Gamma\rho_{23}(\tau) & \Gamma\rho_{24}(\tau) & \Gamma\rho_{25}(\tau) & 3\Gamma\rho_{26}(\tau) & \Gamma\rho_{27}(\tau) & \Gamma\rho_{28}(\tau) & 0 \\
\Gamma\rho_{31}(\tau) & \Gamma\rho_{32}(\tau) & 0 & 3\Gamma\rho_{34}(\tau) & \Gamma\rho_{35}(\tau) & \Gamma\rho_{36}(\tau) & \Gamma\rho_{37}(\tau) & 0 & \Gamma\rho_{39}(\tau) \\
4\Gamma\rho_{41}(\tau) & \Gamma\rho_{42}(\tau) & 3\Gamma\rho_{43}(\tau) & 0 & \Gamma\rho_{45}(\tau) & 4\Gamma\rho_{46}(\tau) & \Gamma\rho_{47}(\tau) & 3\Gamma\rho_{48}(\tau) & \Gamma\rho_{49}(\tau) \\
3\Gamma\rho_{51}(\tau) & \Gamma\rho_{52}(\tau) & \Gamma\rho_{53}(\tau) & \Gamma\rho_{54}(\tau) & 0 & \Gamma\rho_{56}(\tau) & 0 & \Gamma\rho_{58}(\tau) & \Gamma\rho_{59}(\tau) \\
4\Gamma\rho_{61}(\tau) & 3\Gamma\rho_{62}(\tau) & \Gamma\rho_{63}(\tau) & 4\Gamma\rho_{64}(\tau) & \Gamma\rho_{65}(\tau) & 0 & \Gamma\rho_{67}(\tau) & \Gamma\rho_{68}(\tau) & 3\Gamma\rho_{69}(\tau) \\
3\Gamma\rho_{71}(\tau) & \Gamma\rho_{72}(\tau) & \Gamma\rho_{73}(\tau) & \Gamma\rho_{74}(\tau) & 0 & \Gamma\rho_{76}(\tau) & 0 & \Gamma\rho_{78}(\tau) & \Gamma\rho_{79}(\tau) \\
\Gamma\rho_{81}(\tau) & \Gamma\rho_{82}(\tau) & 0 & 3\Gamma\rho_{84}(\tau) & \Gamma\rho_{85}(\tau) & \Gamma\rho_{86}(\tau) & \Gamma\rho_{87}(\tau) & 0 & \Gamma\rho_{89}(\tau) \\
\Gamma\rho_{91}(\tau) & 0 & \Gamma\rho_{93}(\tau) & \Gamma\rho_{94}(\tau) & \Gamma\rho_{95}(\tau) & 3\Gamma\rho_{96}(\tau) & \Gamma\rho_{97}(\tau) & \Gamma\rho_{98}(\tau) & 0 \\
\end{pmatrix}, \\
D_S &= 
\begin{pmatrix}
0 & 0 & 0 & 0 & 0 & 0 & 0 & 0 & 0 \\
0 & 0 & \gamma_{23}(\tau)\rho_{23}(\tau) & 0 & \gamma_{25}(\tau)\rho_{25}(\tau) & 0 & \gamma_{27}(\tau)\rho_{27}(\tau) & \gamma_{28}(\tau)\rho_{28}(\tau) & \gamma_{29}(\tau)\rho_{29}(\tau) \\
0 & \gamma_{32}(\tau)\rho_{32}(\tau) & 0 & 0 & \gamma_{35}(\tau)\rho_{35}(\tau) & 0 & \gamma_{37}(\tau)\rho_{37}(\tau) & \gamma_{38}(\tau)\rho_{38}(\tau) & \gamma_{39}(\tau)\rho_{39}(\tau) \\
0 & 0 & 0 & 0 & 0 & 0 & 0 & 0 & 0 \\
0 & \gamma_{52}(\tau)\rho_{52}(\tau) & \gamma_{53}(\tau)\rho_{53}(\tau) & 0 & 0 & 0 & \gamma_{57}(\tau)\rho_{57}(\tau) & \gamma_{58}(\tau)\rho_{58}(\tau) & \gamma_{59}(\tau)\rho_{59}(\tau) \\
0 & 0 & 0 & 0 & 0 & 0 & 0 & 0 & 0 \\
0 & \gamma_{72}(\tau)\rho_{72}(\tau) & \gamma_{73}(\tau)\rho_{73}(\tau) & 0 & \gamma_{75}(\tau)\rho_{75}(\tau) & 0 & 0 & \gamma_{78}(\tau)\rho_{78}(\tau) & \gamma_{79}(\tau)\rho_{79}(\tau) \\
0 & \gamma_{82}(\tau)\rho_{82}(\tau) & \gamma_{83}(\tau)\rho_{83}(\tau) & 0 & \gamma_{85}(\tau)\rho_{85}(\tau) & 0 & \gamma_{87}(\tau)\rho_{87}(\tau) & 0 & \gamma_{89}(\tau)\rho_{89}(\tau) \\
0 & \gamma_{92}(\tau)\rho_{92}(\tau) & \gamma_{93}(\tau)\rho_{93}(\tau) & 0 & \gamma_{95}(\tau)\rho_{95}(\tau) & 0 & \gamma_{97}(\tau)\rho_{97}(\tau) & \gamma_{98}(\tau)\rho_{98}(\tau) & 0 \\
\end{pmatrix}.
\end{split}
\end{equation*}
\end{widetext}

% Feasibility of the detuning and tunnelling parameters
\section{Feasibility Check on Pulse-Gating Constraints}
\label{appendix:feasible}

%%%%%%%%%%%%%%%%%%%%%%%%%%
\begin{figure}
	\includegraphics[width=3.4in]{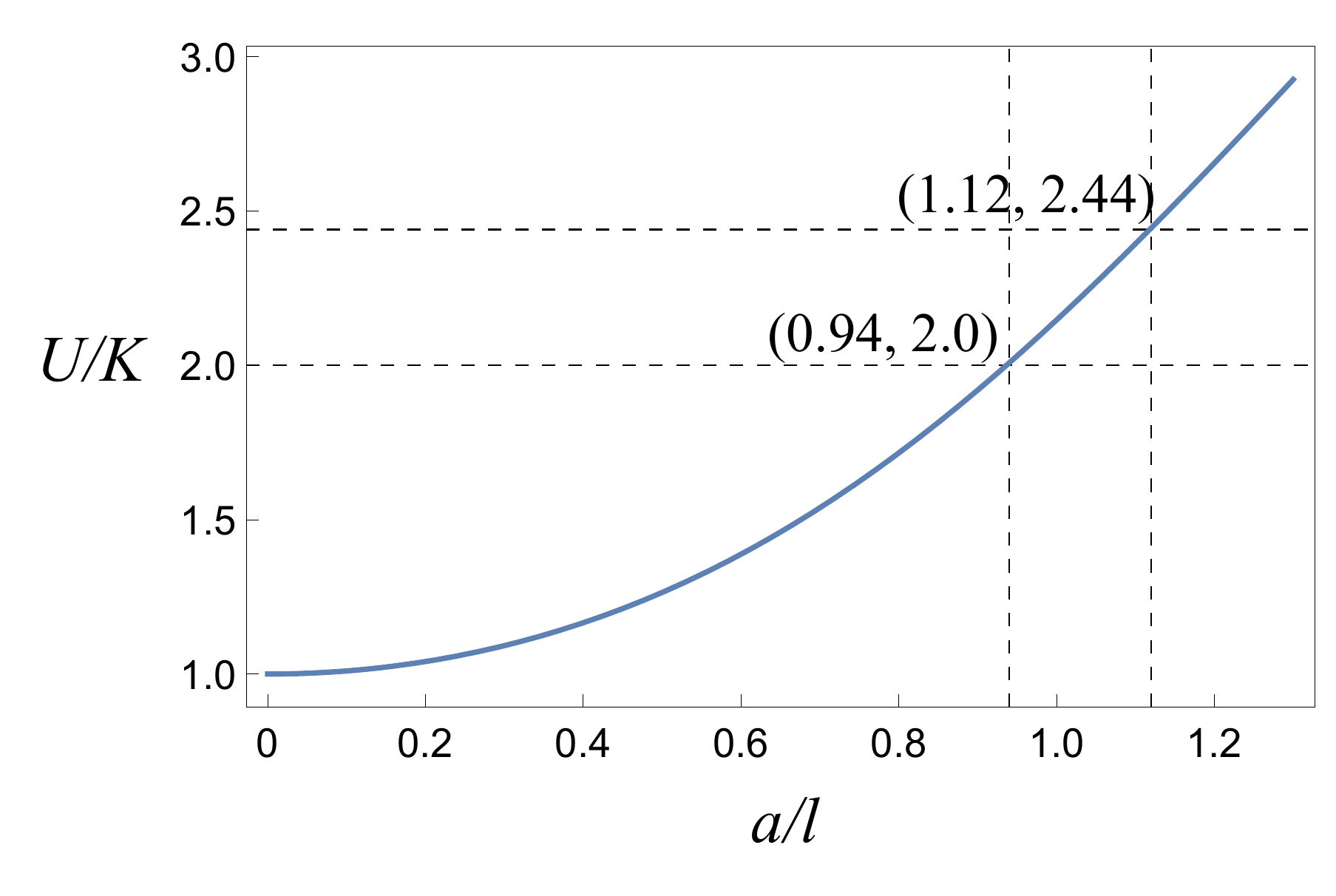}
	\caption{Ratio of intra- to inter-dot Coulomb energy $U/K$ plotted against inter-dot distance to effective dot size ratio $a/l$, for a double dot system modelled by a bi-quadratic potential. The pulse-gated requirement of $U/K=2$ corresponds to an $a/l \approx 0.94$ which is a realistic constraint on the quantum dots. $U/K \approx 2.44$ corresponds to $a/l \approx 1.12$  for the parameters used in the adiabatic scheme.
	}
	\label{fig:besselfn}
\end{figure}
%%%%%%%%%%%%%%%%%%%%%%%%%

The pulse-gated scheme required a condition on the ratio of intra- to inter-dot Coulomb energies, $U/K=2$. We show here, that this condition corresponds to a feasible inter-dot distance to effective dot size ratio, $a/l$. By considering a biquadratic potential for a double quantum dot, where the interdot distance is $2a$ and $l$ is the effective length of the ground, $s$ orbital wavefunction, the intra- and inter-dot Coulomb energies, are respectively given by~\cite{DasSarma:2011p235314}
\begin{align}
U &= ke^2 \sqrt{\frac{\pi}{2}}\frac{1}{l}, \\
K &= ke^2 \sqrt{\frac{\pi}{2l^2}}\exp[-\frac{a^2}{l^2}]I_0\left(\frac{a^2}{l^2}\right),
\end{align}
where $k=1/(4\pi\epsilon_0\epsilon_\text{r})$, $e$ is the electronic charge, $l$ is the width of the dot, $2a$ is the distance between the two dots, $I_0$ is the zeroth order modified Bessel function, $\varepsilon_0$ is the vacuum permittivity ($8.85\times10^{-12}\text{ F m}^{-1}$), and $\varepsilon_\text{r}$ is the dielectric constant (12.375 for Si/SiGe~\cite{DasSarma:2011p235314}). From here we can determine the ratio of $U/K$ to be given by:
\begin{equation*}
\frac{U}{K} = \frac{\exp(\frac{a^2}{l^2})}{I_0(\frac{a^2}{l^2})}
\end{equation*}

The result is that $U/K= 2$ corresponds to an $a/l \approx 0.94$. On the other hand we are not constrained by the $U/K$ ratio in the adiabatic scheme, and the numerical values used yield ratios of $U/K \approx 2.44$ and $a/l \approx 1.12$. A graphical representation of the relation between $U/K$ and $a/l$ is shown in Fig. \ref{fig:besselfn}. As the ratio $a/l$ decreases,  the tunnel barrier becomes thinner relative to the width of the potential well and conversely, as $a/l$ increases, the tunnel barrier  widens relative to the width of the potential wells.

Therefore, in order to engineer the desired $U/K$ ratio, it is sufficient to manipulate the inter-dot distance to dot confinement. This is necessary in the pulse-gated scheme, and is also advantageous in the adiabatic scheme because it allows tunability of the positions of energy gaps via the energies of the singly and doubly occupied states.

\bibliographystyle{apsrev4-1}
\bibliography{MK_Feng_Citations}

\begin{thebibliography}{51}%
\makeatletter
\providecommand \@ifxundefined [1]{%
 \@ifx{#1\undefined}
}%
\providecommand \@ifnum [1]{%
 \ifnum #1\expandafter \@firstoftwo
 \else \expandafter \@secondoftwo
 \fi
}%
\providecommand \@ifx [1]{%
 \ifx #1\expandafter \@firstoftwo
 \else \expandafter \@secondoftwo
 \fi
}%
\providecommand \natexlab [1]{#1}%
\providecommand \enquote  [1]{``#1''}%
\providecommand \bibnamefont  [1]{#1}%
\providecommand \bibfnamefont [1]{#1}%
\providecommand \citenamefont [1]{#1}%
\providecommand \href@noop [0]{\@secondoftwo}%
\providecommand \href [0]{\begingroup \@sanitize@url \@href}%
\providecommand \@href[1]{\@@startlink{#1}\@@href}%
\providecommand \@@href[1]{\endgroup#1\@@endlink}%
\providecommand \@sanitize@url [0]{\catcode `\\12\catcode `\$12\catcode
  `\&12\catcode `\#12\catcode `\^12\catcode `\_12\catcode `\%12\relax}%
\providecommand \@@startlink[1]{}%
\providecommand \@@endlink[0]{}%
\providecommand \url  [0]{\begingroup\@sanitize@url \@url }%
\providecommand \@url [1]{\endgroup\@href {#1}{\urlprefix }}%
\providecommand \urlprefix  [0]{URL }%
\providecommand \Eprint [0]{\href }%
\providecommand \doibase [0]{http://dx.doi.org/}%
\providecommand \selectlanguage [0]{\@gobble}%
\providecommand \bibinfo  [0]{\@secondoftwo}%
\providecommand \bibfield  [0]{\@secondoftwo}%
\providecommand \translation [1]{[#1]}%
\providecommand \BibitemOpen [0]{}%
\providecommand \bibitemStop [0]{}%
\providecommand \bibitemNoStop [0]{.\EOS\space}%
\providecommand \EOS [0]{\spacefactor3000\relax}%
\providecommand \BibitemShut  [1]{\csname bibitem#1\endcsname}%
\let\auto@bib@innerbib\@empty
%</preamble>
\bibitem [{\citenamefont {Koppens}\ \emph {et~al.}(2006)\citenamefont
  {Koppens}, \citenamefont {Buizert}, \citenamefont {Tielrooij}, \citenamefont
  {Vink}, \citenamefont {Nowack}, \citenamefont {Meunier}, \citenamefont
  {Kouwenhoven},\ and\ \citenamefont {Vandersypen}}]{Koppens:2006p766}%
  \BibitemOpen
  \bibfield  {author} {\bibinfo {author} {\bibfnamefont {F.~H.~L.}\
  \bibnamefont {Koppens}}, \bibinfo {author} {\bibfnamefont {C.}~\bibnamefont
  {Buizert}}, \bibinfo {author} {\bibfnamefont {K.~J.}\ \bibnamefont
  {Tielrooij}}, \bibinfo {author} {\bibfnamefont {I.~T.}\ \bibnamefont {Vink}},
  \bibinfo {author} {\bibfnamefont {K.~C.}\ \bibnamefont {Nowack}}, \bibinfo
  {author} {\bibfnamefont {T.}~\bibnamefont {Meunier}}, \bibinfo {author}
  {\bibfnamefont {L.~P.}\ \bibnamefont {Kouwenhoven}}, \ and\ \bibinfo {author}
  {\bibfnamefont {L.~M.~K.}\ \bibnamefont {Vandersypen}},\ }\href {\doibase
  10.1038/nature05065} {\bibfield  {journal} {\bibinfo  {journal} {Nature}\
  }\textbf {\bibinfo {volume} {442}},\ \bibinfo {pages} {766} (\bibinfo {year}
  {2006})}\BibitemShut {NoStop}%
\bibitem [{\citenamefont {Bluhm}\ \emph {et~al.}(2011)\citenamefont {Bluhm},
  \citenamefont {Foletti}, \citenamefont {Neder}, \citenamefont {Rudner},
  \citenamefont {Mahalu}, \citenamefont {Umansky},\ and\ \citenamefont
  {Yacoby}}]{Bluhm:2011p109}%
  \BibitemOpen
  \bibfield  {author} {\bibinfo {author} {\bibfnamefont {H.}~\bibnamefont
  {Bluhm}}, \bibinfo {author} {\bibfnamefont {S.}~\bibnamefont {Foletti}},
  \bibinfo {author} {\bibfnamefont {I.}~\bibnamefont {Neder}}, \bibinfo
  {author} {\bibfnamefont {M.}~\bibnamefont {Rudner}}, \bibinfo {author}
  {\bibfnamefont {D.}~\bibnamefont {Mahalu}}, \bibinfo {author} {\bibfnamefont
  {V.}~\bibnamefont {Umansky}}, \ and\ \bibinfo {author} {\bibfnamefont
  {A.}~\bibnamefont {Yacoby}},\ }\href {\doibase 10.1038/nphys1856} {\bibfield
  {journal} {\bibinfo  {journal} {Nature Physics}\ }\textbf {\bibinfo {volume}
  {7}},\ \bibinfo {pages} {109} (\bibinfo {year} {2011})}\BibitemShut {NoStop}%
\bibitem [{\citenamefont {Loss}\ and\ \citenamefont
  {DiVincenzo}(1998)}]{Loss:1998p120}%
  \BibitemOpen
  \bibfield  {author} {\bibinfo {author} {\bibfnamefont {D.}~\bibnamefont
  {Loss}}\ and\ \bibinfo {author} {\bibfnamefont {D.~P.}\ \bibnamefont
  {DiVincenzo}},\ }\href {\doibase 10.1103/PhysRevA.57.120} {\bibfield
  {journal} {\bibinfo  {journal} {Phys.~Rev.~A}\ }\textbf {\bibinfo {volume}
  {57}},\ \bibinfo {pages} {120} (\bibinfo {year} {1998})}\BibitemShut
  {NoStop}%
\bibitem [{\citenamefont {Levy}(2002)}]{Levy:2002p147902}%
  \BibitemOpen
  \bibfield  {author} {\bibinfo {author} {\bibfnamefont {J.}~\bibnamefont
  {Levy}},\ }\href {\doibase 10.1103/PhysRevLett.89.147902} {\bibfield
  {journal} {\bibinfo  {journal} {Phys.~Rev.~Lett.}\ }\textbf {\bibinfo
  {volume} {89}},\ \bibinfo {pages} {147902} (\bibinfo {year}
  {2002})}\BibitemShut {NoStop}%
\bibitem [{\citenamefont {Petta}\ \emph {et~al.}(2010)\citenamefont {Petta},
  \citenamefont {Lu},\ and\ \citenamefont {Gossard}}]{Petta:2010p669}%
  \BibitemOpen
  \bibfield  {author} {\bibinfo {author} {\bibfnamefont {J.~R.}\ \bibnamefont
  {Petta}}, \bibinfo {author} {\bibfnamefont {H.}~\bibnamefont {Lu}}, \ and\
  \bibinfo {author} {\bibfnamefont {A.~C.}\ \bibnamefont {Gossard}},\ }\href
  {\doibase 10.1126/science.1183628} {\bibfield  {journal} {\bibinfo  {journal}
  {Science}\ }\textbf {\bibinfo {volume} {327}},\ \bibinfo {pages} {669}
  (\bibinfo {year} {2010})}\BibitemShut {NoStop}%
\bibitem [{\citenamefont {DiVincenzo}\ \emph {et~al.}(2000)\citenamefont
  {DiVincenzo}, \citenamefont {Bacon}, \citenamefont {Kempe}, \citenamefont
  {Burkard},\ and\ \citenamefont {Whaley}}]{DiVincenzo:2000p339}%
  \BibitemOpen
  \bibfield  {author} {\bibinfo {author} {\bibfnamefont {D.~P.}\ \bibnamefont
  {DiVincenzo}}, \bibinfo {author} {\bibfnamefont {D.}~\bibnamefont {Bacon}},
  \bibinfo {author} {\bibfnamefont {J.}~\bibnamefont {Kempe}}, \bibinfo
  {author} {\bibfnamefont {G.}~\bibnamefont {Burkard}}, \ and\ \bibinfo
  {author} {\bibfnamefont {K.~B.}\ \bibnamefont {Whaley}},\ }\href {\doibase
  10.1038/35042541} {\bibfield  {journal} {\bibinfo  {journal} {Nature}\
  }\textbf {\bibinfo {volume} {408}},\ \bibinfo {pages} {339} (\bibinfo {year}
  {2000})}\BibitemShut {NoStop}%
\bibitem [{\citenamefont {Laird}\ \emph {et~al.}(2010)\citenamefont {Laird},
  \citenamefont {Taylor}, \citenamefont {DiVincenzo}, \citenamefont {Marcus},
  \citenamefont {Hanson},\ and\ \citenamefont {Gossard}}]{Laird:2010p075403}%
  \BibitemOpen
  \bibfield  {author} {\bibinfo {author} {\bibfnamefont {E.~A.}\ \bibnamefont
  {Laird}}, \bibinfo {author} {\bibfnamefont {J.~M.}\ \bibnamefont {Taylor}},
  \bibinfo {author} {\bibfnamefont {D.~P.}\ \bibnamefont {DiVincenzo}},
  \bibinfo {author} {\bibfnamefont {C.~M.}\ \bibnamefont {Marcus}}, \bibinfo
  {author} {\bibfnamefont {M.~P.}\ \bibnamefont {Hanson}}, \ and\ \bibinfo
  {author} {\bibfnamefont {A.~C.}\ \bibnamefont {Gossard}},\ }\href {\doibase
  10.1103/PhysRevB.82.075403} {\bibfield  {journal} {\bibinfo  {journal}
  {Phys.~Rev.~B}\ }\textbf {\bibinfo {volume} {82}},\ \bibinfo {pages} {075403}
  (\bibinfo {year} {2010})}\BibitemShut {NoStop}%
\bibitem [{\citenamefont {Svore}\ \emph {et~al.}(2005)\citenamefont {Svore},
  \citenamefont {Terhal},\ and\ \citenamefont
  {DiVincenzo}}]{Svore:2005p022317}%
  \BibitemOpen
  \bibfield  {author} {\bibinfo {author} {\bibfnamefont {K.~M.}\ \bibnamefont
  {Svore}}, \bibinfo {author} {\bibfnamefont {B.~M.}\ \bibnamefont {Terhal}}, \
  and\ \bibinfo {author} {\bibfnamefont {D.~P.}\ \bibnamefont {DiVincenzo}},\
  }\href {\doibase 10.1103/PhysRevA.72.022317} {\bibfield  {journal} {\bibinfo
  {journal} {Phys. Rev. A}\ }\textbf {\bibinfo {volume} {72}},\ \bibinfo
  {pages} {022317} (\bibinfo {year} {2005})}\BibitemShut {NoStop}%
\bibitem [{\citenamefont {Li}\ \emph {et~al.}(2010)\citenamefont {Li},
  \citenamefont {Cywinski}, \citenamefont {Culcer}, \citenamefont {Hu},\ and\
  \citenamefont {{Das Sarma}}}]{Li:2010p085313}%
  \BibitemOpen
  \bibfield  {author} {\bibinfo {author} {\bibfnamefont {Q.}~\bibnamefont
  {Li}}, \bibinfo {author} {\bibfnamefont {L.}~\bibnamefont {Cywinski}},
  \bibinfo {author} {\bibfnamefont {D.}~\bibnamefont {Culcer}}, \bibinfo
  {author} {\bibfnamefont {X.}~\bibnamefont {Hu}}, \ and\ \bibinfo {author}
  {\bibfnamefont {S.}~\bibnamefont {{Das Sarma}}},\ }\href {\doibase
  10.1103/PhysRevB.81.085313} {\bibfield  {journal} {\bibinfo  {journal} {Phys.
  Rev. B}\ }\textbf {\bibinfo {volume} {81}},\ \bibinfo {pages} {085313}
  (\bibinfo {year} {2010})}\BibitemShut {NoStop}%
\bibitem [{\citenamefont {Taylor}\ \emph {et~al.}(2005)\citenamefont {Taylor},
  \citenamefont {Engel}, \citenamefont {D$\ddot{\text{u}}$r}, \citenamefont
  {Yacoby}, \citenamefont {Marcus}, \citenamefont {Zoller},\ and\ \citenamefont
  {Lukin}}]{Taylor:2005p177}%
  \BibitemOpen
  \bibfield  {author} {\bibinfo {author} {\bibfnamefont {J.~M.}\ \bibnamefont
  {Taylor}}, \bibinfo {author} {\bibfnamefont {H.-A.}\ \bibnamefont {Engel}},
  \bibinfo {author} {\bibfnamefont {W.}~\bibnamefont {D$\ddot{\text{u}}$r}},
  \bibinfo {author} {\bibfnamefont {A.}~\bibnamefont {Yacoby}}, \bibinfo
  {author} {\bibfnamefont {C.~M.}\ \bibnamefont {Marcus}}, \bibinfo {author}
  {\bibfnamefont {P.}~\bibnamefont {Zoller}}, \ and\ \bibinfo {author}
  {\bibfnamefont {M.~D.}\ \bibnamefont {Lukin}},\ }\href {\doibase
  10.1038/nphys174} {\bibfield  {journal} {\bibinfo  {journal} {Nature
  Physics}\ }\textbf {\bibinfo {volume} {1}},\ \bibinfo {pages} {177} (\bibinfo
  {year} {2005})}\BibitemShut {NoStop}%
\bibitem [{\citenamefont {Flentje}\ \emph {et~al.}(2017)\citenamefont
  {Flentje}, \citenamefont {Mortemousque}, \citenamefont {Thalineau},
  \citenamefont {Ludwig}, \citenamefont {Wieck}, \citenamefont {B{\"a}uerle},\
  and\ \citenamefont {Meunier}}]{flentje_2017}%
  \BibitemOpen
  \bibfield  {author} {\bibinfo {author} {\bibfnamefont {H.}~\bibnamefont
  {Flentje}}, \bibinfo {author} {\bibfnamefont {P.-A.}\ \bibnamefont
  {Mortemousque}}, \bibinfo {author} {\bibfnamefont {R.}~\bibnamefont
  {Thalineau}}, \bibinfo {author} {\bibfnamefont {A.}~\bibnamefont {Ludwig}},
  \bibinfo {author} {\bibfnamefont {A.~D.}\ \bibnamefont {Wieck}}, \bibinfo
  {author} {\bibfnamefont {C.}~\bibnamefont {B{\"a}uerle}}, \ and\ \bibinfo
  {author} {\bibfnamefont {T.}~\bibnamefont {Meunier}},\ }\href {\doibase
  10.1038/s41467-017-00534-3} {\bibfield  {journal} {\bibinfo  {journal}
  {Nature Communications}\ }\textbf {\bibinfo {volume} {8}} (\bibinfo {year}
  {2017}),\ 10.1038/s41467-017-00534-3}\BibitemShut {NoStop}%
\bibitem [{\citenamefont {Fujita}\ \emph {et~al.}(2017)\citenamefont {Fujita},
  \citenamefont {Baart}, \citenamefont {Reichl}, \citenamefont {Wegscheider},\
  and\ \citenamefont {Vandersypen}}]{fujita_2017}%
  \BibitemOpen
  \bibfield  {author} {\bibinfo {author} {\bibfnamefont {T.}~\bibnamefont
  {Fujita}}, \bibinfo {author} {\bibfnamefont {T.~A.}\ \bibnamefont {Baart}},
  \bibinfo {author} {\bibfnamefont {C.}~\bibnamefont {Reichl}}, \bibinfo
  {author} {\bibfnamefont {W.}~\bibnamefont {Wegscheider}}, \ and\ \bibinfo
  {author} {\bibfnamefont {L.~M.~K.}\ \bibnamefont {Vandersypen}},\ }\href
  {\doibase 10.1038/s41534-017-0024-4} {\bibfield  {journal} {\bibinfo
  {journal} {npj Quantum Information}\ }\textbf {\bibinfo {volume} {3}}
  (\bibinfo {year} {2017}),\ 10.1038/s41534-017-0024-4}\BibitemShut {NoStop}%
\bibitem [{\citenamefont {Christandl}\ \emph {et~al.}(2004)\citenamefont
  {Christandl}, \citenamefont {Datta}, \citenamefont {Ekert},\ and\
  \citenamefont {Landahl}}]{Christandl:2004p187902}%
  \BibitemOpen
  \bibfield  {author} {\bibinfo {author} {\bibfnamefont {M.}~\bibnamefont
  {Christandl}}, \bibinfo {author} {\bibfnamefont {N.}~\bibnamefont {Datta}},
  \bibinfo {author} {\bibfnamefont {A.}~\bibnamefont {Ekert}}, \ and\ \bibinfo
  {author} {\bibfnamefont {A.~J.}\ \bibnamefont {Landahl}},\ }\href {\doibase
  10.1103/PhysRevLett.92.187902} {\bibfield  {journal} {\bibinfo  {journal}
  {Phys. Rev. Lett.}\ }\textbf {\bibinfo {volume} {92}},\ \bibinfo {pages}
  {187902} (\bibinfo {year} {2004})}\BibitemShut {NoStop}%
\bibitem [{\citenamefont {Friesen}\ \emph {et~al.}(2007)\citenamefont
  {Friesen}, \citenamefont {Biswas}, \citenamefont {Hu},\ and\ \citenamefont
  {Lidar}}]{Friesen:2007p230503}%
  \BibitemOpen
  \bibfield  {author} {\bibinfo {author} {\bibfnamefont {M.}~\bibnamefont
  {Friesen}}, \bibinfo {author} {\bibfnamefont {A.}~\bibnamefont {Biswas}},
  \bibinfo {author} {\bibfnamefont {X.}~\bibnamefont {Hu}}, \ and\ \bibinfo
  {author} {\bibfnamefont {D.}~\bibnamefont {Lidar}},\ }\href {\doibase
  10.1103/PhysRevLett.98.230503} {\bibfield  {journal} {\bibinfo  {journal}
  {Phys. Rev. Lett.}\ }\textbf {\bibinfo {volume} {98}},\ \bibinfo {pages}
  {230503} (\bibinfo {year} {2007})}\BibitemShut {NoStop}%
\bibitem [{\citenamefont {Greentree}\ \emph {et~al.}(2004)\citenamefont
  {Greentree}, \citenamefont {Cole}, \citenamefont {Hamilton},\ and\
  \citenamefont {Hollenberg}}]{Greentree:2004p235317}%
  \BibitemOpen
  \bibfield  {author} {\bibinfo {author} {\bibfnamefont {A.~D.}\ \bibnamefont
  {Greentree}}, \bibinfo {author} {\bibfnamefont {J.~H.}\ \bibnamefont {Cole}},
  \bibinfo {author} {\bibfnamefont {A.~R.}\ \bibnamefont {Hamilton}}, \ and\
  \bibinfo {author} {\bibfnamefont {L.~C.~L.}\ \bibnamefont {Hollenberg}},\
  }\href {\doibase https://doi.org/10.1103/PhysRevB.70.235317} {\bibfield
  {journal} {\bibinfo  {journal} {Phys. Rev. B}\ }\textbf {\bibinfo {volume}
  {70}},\ \bibinfo {pages} {235317} (\bibinfo {year} {2004})}\BibitemShut
  {NoStop}%
\bibitem [{\citenamefont {Ferraro}\ \emph {et~al.}(2015)\citenamefont
  {Ferraro}, \citenamefont {DeMichielis}, \citenamefont {Fanciulli},\ and\
  \citenamefont {Prati}}]{Ferraro:2015p075435}%
  \BibitemOpen
  \bibfield  {author} {\bibinfo {author} {\bibfnamefont {E.}~\bibnamefont
  {Ferraro}}, \bibinfo {author} {\bibfnamefont {M.}~\bibnamefont
  {DeMichielis}}, \bibinfo {author} {\bibfnamefont {M.}~\bibnamefont
  {Fanciulli}}, \ and\ \bibinfo {author} {\bibfnamefont {E.}~\bibnamefont
  {Prati}},\ }\href {\doibase 10.1103/PhysRevB.91.075435} {\bibfield  {journal}
  {\bibinfo  {journal} {Phys. Rev. B}\ }\textbf {\bibinfo {volume} {91}},\
  \bibinfo {pages} {075435} (\bibinfo {year} {2015})}\BibitemShut {NoStop}%
\bibitem [{\citenamefont {Imamoglu}\ \emph {et~al.}(1999)\citenamefont
  {Imamoglu}, \citenamefont {Awschalom}, \citenamefont {Burkard}, \citenamefont
  {DiVincenzo}, \citenamefont {Loss}, \citenamefont {Sherwin},\ and\
  \citenamefont {Small}}]{Imamoglu:1999p4204}%
  \BibitemOpen
  \bibfield  {author} {\bibinfo {author} {\bibfnamefont {A.}~\bibnamefont
  {Imamoglu}}, \bibinfo {author} {\bibfnamefont {D.~D.}\ \bibnamefont
  {Awschalom}}, \bibinfo {author} {\bibfnamefont {G.}~\bibnamefont {Burkard}},
  \bibinfo {author} {\bibfnamefont {D.~P.}\ \bibnamefont {DiVincenzo}},
  \bibinfo {author} {\bibfnamefont {D.}~\bibnamefont {Loss}}, \bibinfo {author}
  {\bibfnamefont {M.}~\bibnamefont {Sherwin}}, \ and\ \bibinfo {author}
  {\bibfnamefont {A.}~\bibnamefont {Small}},\ }\href {\doibase
  10.1103/PhysRevLett.83.4204} {\bibfield  {journal} {\bibinfo  {journal}
  {Phys. Rev. Lett.}\ }\textbf {\bibinfo {volume} {83}},\ \bibinfo {pages}
  {4204} (\bibinfo {year} {1999})}\BibitemShut {NoStop}%
\bibitem [{\citenamefont {Hu}\ \emph {et~al.}(2012)\citenamefont {Hu},
  \citenamefont {Liu},\ and\ \citenamefont {Nori}}]{Hu:2012p035314}%
  \BibitemOpen
  \bibfield  {author} {\bibinfo {author} {\bibfnamefont {X.}~\bibnamefont
  {Hu}}, \bibinfo {author} {\bibfnamefont {Y.-X.}\ \bibnamefont {Liu}}, \ and\
  \bibinfo {author} {\bibfnamefont {F.}~\bibnamefont {Nori}},\ }\href {\doibase
  10.1103/PhysRevB.86.035314} {\bibfield  {journal} {\bibinfo  {journal} {Phys.
  Rev. B}\ }\textbf {\bibinfo {volume} {86}},\ \bibinfo {pages} {035314}
  (\bibinfo {year} {2012})}\BibitemShut {NoStop}%
\bibitem [{\citenamefont {Petersson}\ \emph {et~al.}(2012)\citenamefont
  {Petersson}, \citenamefont {McFaul}, \citenamefont {Schroer}, \citenamefont
  {Jung}, \citenamefont {Taylor}, \citenamefont {Houck},\ and\ \citenamefont
  {Petta}}]{Petersson:2012p380}%
  \BibitemOpen
  \bibfield  {author} {\bibinfo {author} {\bibfnamefont {K.~D.}\ \bibnamefont
  {Petersson}}, \bibinfo {author} {\bibfnamefont {L.~W.}\ \bibnamefont
  {McFaul}}, \bibinfo {author} {\bibfnamefont {M.~D.}\ \bibnamefont {Schroer}},
  \bibinfo {author} {\bibfnamefont {M.}~\bibnamefont {Jung}}, \bibinfo {author}
  {\bibfnamefont {J.~M.}\ \bibnamefont {Taylor}}, \bibinfo {author}
  {\bibfnamefont {A.~A.}\ \bibnamefont {Houck}}, \ and\ \bibinfo {author}
  {\bibfnamefont {J.~R.}\ \bibnamefont {Petta}},\ }\href {\doibase
  doi:10.1038/nature11559} {\bibfield  {journal} {\bibinfo  {journal} {Nature}\
  }\textbf {\bibinfo {volume} {490}},\ \bibinfo {pages} {380} (\bibinfo {year}
  {2012})}\BibitemShut {NoStop}%
\bibitem [{\citenamefont {Vermersch}\ \emph {et~al.}(2017)\citenamefont
  {Vermersch}, \citenamefont {Guimond}, \citenamefont {Pichler},\ and\
  \citenamefont {Zoller}}]{Vermersch_2017}%
  \BibitemOpen
  \bibfield  {author} {\bibinfo {author} {\bibfnamefont {B.}~\bibnamefont
  {Vermersch}}, \bibinfo {author} {\bibfnamefont {P.-O.}\ \bibnamefont
  {Guimond}}, \bibinfo {author} {\bibfnamefont {H.}~\bibnamefont {Pichler}}, \
  and\ \bibinfo {author} {\bibfnamefont {P.}~\bibnamefont {Zoller}},\ }\href
  {\doibase 10.1103/PhysRevLett.118.133601} {\bibfield  {journal} {\bibinfo
  {journal} {Phys. Rev. Lett.}\ }\textbf {\bibinfo {volume} {118}},\ \bibinfo
  {pages} {133601} (\bibinfo {year} {2017})}\BibitemShut {NoStop}%
\bibitem [{\citenamefont {Wu}\ \emph {et~al.}(2014)\citenamefont {Wu},
  \citenamefont {Ward}, \citenamefont {Prance}, \citenamefont {Kim},
  \citenamefont {Gamble}, \citenamefont {Mohr}, \citenamefont {Shi},
  \citenamefont {Savage}, \citenamefont {Lagally}, \citenamefont {Friesen},
  \citenamefont {Coppersmith},\ and\ \citenamefont {Eriksson}}]{Wu:2014p11938}%
  \BibitemOpen
  \bibfield  {author} {\bibinfo {author} {\bibfnamefont {X.}~\bibnamefont
  {Wu}}, \bibinfo {author} {\bibfnamefont {D.~R.}\ \bibnamefont {Ward}},
  \bibinfo {author} {\bibfnamefont {J.~R.}\ \bibnamefont {Prance}}, \bibinfo
  {author} {\bibfnamefont {D.}~\bibnamefont {Kim}}, \bibinfo {author}
  {\bibfnamefont {J.}~\bibnamefont {Gamble}}, \bibinfo {author} {\bibfnamefont
  {R.~T.}\ \bibnamefont {Mohr}}, \bibinfo {author} {\bibfnamefont
  {Z.}~\bibnamefont {Shi}}, \bibinfo {author} {\bibfnamefont {D.~E.}\
  \bibnamefont {Savage}}, \bibinfo {author} {\bibfnamefont {M.~G.}\
  \bibnamefont {Lagally}}, \bibinfo {author} {\bibfnamefont {M.}~\bibnamefont
  {Friesen}}, \bibinfo {author} {\bibfnamefont {S.~N.}\ \bibnamefont
  {Coppersmith}}, \ and\ \bibinfo {author} {\bibfnamefont {M.~A.}\ \bibnamefont
  {Eriksson}},\ }\href {\doibase 10.1073/pnas.1412230111} {\bibfield  {journal}
  {\bibinfo  {journal} {Proc. Nat. Acad. Sci.}\ }\textbf {\bibinfo {volume}
  {111}},\ \bibinfo {pages} {11938} (\bibinfo {year} {2014})}\BibitemShut
  {NoStop}%
\bibitem [{\citenamefont {Shulman}\ \emph {et~al.}(2012)\citenamefont
  {Shulman}, \citenamefont {Dial}, \citenamefont {Harvey}, \citenamefont
  {Bluhm}, \citenamefont {Umansky},\ and\ \citenamefont
  {Yacoby}}]{Shulman:2012p202}%
  \BibitemOpen
  \bibfield  {author} {\bibinfo {author} {\bibfnamefont {M.~D.}\ \bibnamefont
  {Shulman}}, \bibinfo {author} {\bibfnamefont {O.~E.}\ \bibnamefont {Dial}},
  \bibinfo {author} {\bibfnamefont {S.~P.}\ \bibnamefont {Harvey}}, \bibinfo
  {author} {\bibfnamefont {H.}~\bibnamefont {Bluhm}}, \bibinfo {author}
  {\bibfnamefont {V.}~\bibnamefont {Umansky}}, \ and\ \bibinfo {author}
  {\bibfnamefont {A.}~\bibnamefont {Yacoby}},\ }\href {\doibase
  10.1126/science.1217692} {\bibfield  {journal} {\bibinfo  {journal}
  {Science}\ }\textbf {\bibinfo {volume} {336}},\ \bibinfo {pages} {202}
  (\bibinfo {year} {2012})}\BibitemShut {NoStop}%
\bibitem [{\citenamefont {Maune}\ \emph {et~al.}(2012)\citenamefont {Maune},
  \citenamefont {Borselli}, \citenamefont {Huang}, \citenamefont {Ladd},
  \citenamefont {Deelman}, \citenamefont {Holabird}, \citenamefont {Kiselev},
  \citenamefont {Alvarado-Rodriguez}, \citenamefont {Ross}, \citenamefont
  {Schmitz}, \citenamefont {Sokolich}, \citenamefont {Watson}, \citenamefont
  {Gyure},\ and\ \citenamefont {Hunter}}]{Maune:2012p344}%
  \BibitemOpen
  \bibfield  {author} {\bibinfo {author} {\bibfnamefont {B.~M.}\ \bibnamefont
  {Maune}}, \bibinfo {author} {\bibfnamefont {M.~G.}\ \bibnamefont {Borselli}},
  \bibinfo {author} {\bibfnamefont {B.}~\bibnamefont {Huang}}, \bibinfo
  {author} {\bibfnamefont {T.~D.}\ \bibnamefont {Ladd}}, \bibinfo {author}
  {\bibfnamefont {P.~W.}\ \bibnamefont {Deelman}}, \bibinfo {author}
  {\bibfnamefont {K.~S.}\ \bibnamefont {Holabird}}, \bibinfo {author}
  {\bibfnamefont {A.~A.}\ \bibnamefont {Kiselev}}, \bibinfo {author}
  {\bibfnamefont {I.}~\bibnamefont {Alvarado-Rodriguez}}, \bibinfo {author}
  {\bibfnamefont {R.~S.}\ \bibnamefont {Ross}}, \bibinfo {author}
  {\bibfnamefont {A.~E.}\ \bibnamefont {Schmitz}}, \bibinfo {author}
  {\bibfnamefont {M.}~\bibnamefont {Sokolich}}, \bibinfo {author}
  {\bibfnamefont {C.~A.}\ \bibnamefont {Watson}}, \bibinfo {author}
  {\bibfnamefont {M.~F.}\ \bibnamefont {Gyure}}, \ and\ \bibinfo {author}
  {\bibfnamefont {A.~T.}\ \bibnamefont {Hunter}},\ }\href {\doibase
  10.1038/nature10707} {\bibfield  {journal} {\bibinfo  {journal} {Nature}\
  }\textbf {\bibinfo {volume} {481}},\ \bibinfo {pages} {344} (\bibinfo {year}
  {2012})}\BibitemShut {NoStop}%
\bibitem [{\citenamefont {Reilly}\ \emph {et~al.}(2008)\citenamefont {Reilly},
  \citenamefont {Taylor}, \citenamefont {Petta}, \citenamefont {Marcus},
  \citenamefont {Hanson},\ and\ \citenamefont {Gossard}}]{Reilly:2008p817}%
  \BibitemOpen
  \bibfield  {author} {\bibinfo {author} {\bibfnamefont {D.~J.}\ \bibnamefont
  {Reilly}}, \bibinfo {author} {\bibfnamefont {J.~M.}\ \bibnamefont {Taylor}},
  \bibinfo {author} {\bibfnamefont {J.~R.}\ \bibnamefont {Petta}}, \bibinfo
  {author} {\bibfnamefont {C.~M.}\ \bibnamefont {Marcus}}, \bibinfo {author}
  {\bibfnamefont {M.~P.}\ \bibnamefont {Hanson}}, \ and\ \bibinfo {author}
  {\bibfnamefont {A.~C.}\ \bibnamefont {Gossard}},\ }\href {\doibase
  10.1126/science.1159221} {\bibfield  {journal} {\bibinfo  {journal}
  {Science}\ }\textbf {\bibinfo {volume} {321}},\ \bibinfo {pages} {817}
  (\bibinfo {year} {2008})}\BibitemShut {NoStop}%
\bibitem [{\citenamefont {Foletti}\ \emph {et~al.}(2009)\citenamefont
  {Foletti}, \citenamefont {Bluhm}, \citenamefont {Mahalu}, \citenamefont
  {Umansky},\ and\ \citenamefont {Yacoby}}]{Foletti:2009p903}%
  \BibitemOpen
  \bibfield  {author} {\bibinfo {author} {\bibfnamefont {S.}~\bibnamefont
  {Foletti}}, \bibinfo {author} {\bibfnamefont {H.}~\bibnamefont {Bluhm}},
  \bibinfo {author} {\bibfnamefont {D.}~\bibnamefont {Mahalu}}, \bibinfo
  {author} {\bibfnamefont {V.}~\bibnamefont {Umansky}}, \ and\ \bibinfo
  {author} {\bibfnamefont {A.}~\bibnamefont {Yacoby}},\ }\href {\doibase
  10.1038/nphys1424} {\bibfield  {journal} {\bibinfo  {journal} {Nature
  Physics}\ }\textbf {\bibinfo {volume} {5}},\ \bibinfo {pages} {903} (\bibinfo
  {year} {2009})}\BibitemShut {NoStop}%
\bibitem [{\citenamefont {Zwanenburg}\ \emph {et~al.}(2013)\citenamefont
  {Zwanenburg}, \citenamefont {Dzurak}, \citenamefont {Morello}, \citenamefont
  {Simmons}, \citenamefont {Hollenberg}, \citenamefont {Klimeck}, \citenamefont
  {Rogge}, \citenamefont {Coppersmith},\ and\ \citenamefont
  {Eriksson}}]{Zwanenburg:2013p961}%
  \BibitemOpen
  \bibfield  {author} {\bibinfo {author} {\bibfnamefont {F.~A.}\ \bibnamefont
  {Zwanenburg}}, \bibinfo {author} {\bibfnamefont {A.~S.}\ \bibnamefont
  {Dzurak}}, \bibinfo {author} {\bibfnamefont {A.}~\bibnamefont {Morello}},
  \bibinfo {author} {\bibfnamefont {M.~Y.}\ \bibnamefont {Simmons}}, \bibinfo
  {author} {\bibfnamefont {L.~C.~L.}\ \bibnamefont {Hollenberg}}, \bibinfo
  {author} {\bibfnamefont {G.}~\bibnamefont {Klimeck}}, \bibinfo {author}
  {\bibfnamefont {S.}~\bibnamefont {Rogge}}, \bibinfo {author} {\bibfnamefont
  {S.~N.}\ \bibnamefont {Coppersmith}}, \ and\ \bibinfo {author} {\bibfnamefont
  {M.~A.}\ \bibnamefont {Eriksson}},\ }\href {\doibase
  10.1103/RevModPhys.85.961} {\bibfield  {journal} {\bibinfo  {journal} {Rev.
  Mod. Phys.}\ }\textbf {\bibinfo {volume} {85}},\ \bibinfo {pages} {961}
  (\bibinfo {year} {2013})}\BibitemShut {NoStop}%
\bibitem [{\citenamefont {Tahan}\ \emph {et~al.}(2002)\citenamefont {Tahan},
  \citenamefont {Friesen},\ and\ \citenamefont {Joynt}}]{Tahan:2002p035314}%
  \BibitemOpen
  \bibfield  {author} {\bibinfo {author} {\bibfnamefont {C.}~\bibnamefont
  {Tahan}}, \bibinfo {author} {\bibfnamefont {M.}~\bibnamefont {Friesen}}, \
  and\ \bibinfo {author} {\bibfnamefont {R.}~\bibnamefont {Joynt}},\ }\href
  {\doibase 10.1103/PhysRevB.66.035314} {\bibfield  {journal} {\bibinfo
  {journal} {Phys. Rev. B}\ }\textbf {\bibinfo {volume} {66}},\ \bibinfo
  {pages} {035314} (\bibinfo {year} {2002})}\BibitemShut {NoStop}%
\bibitem [{\citenamefont {Yang}\ \emph {et~al.}(2011)\citenamefont {Yang},
  \citenamefont {Wang},\ and\ \citenamefont {{Das Sarma}}}]{Yang:2011p161301}%
  \BibitemOpen
  \bibfield  {author} {\bibinfo {author} {\bibfnamefont {S.}~\bibnamefont
  {Yang}}, \bibinfo {author} {\bibfnamefont {X.}~\bibnamefont {Wang}}, \ and\
  \bibinfo {author} {\bibfnamefont {S.}~\bibnamefont {{Das Sarma}}},\ }\href
  {\doibase 10.1103/PhysRevB.83.161301} {\bibfield  {journal} {\bibinfo
  {journal} {Phys. Rev. B}\ }\textbf {\bibinfo {volume} {83}},\ \bibinfo
  {pages} {161301} (\bibinfo {year} {2011})}\BibitemShut {NoStop}%
\bibitem [{\citenamefont {{Das Sarma}}\ \emph {et~al.}(2011)\citenamefont {{Das
  Sarma}}, \citenamefont {Wang},\ and\ \citenamefont
  {Yang}}]{DasSarma:2011p235314}%
  \BibitemOpen
  \bibfield  {author} {\bibinfo {author} {\bibfnamefont {S.}~\bibnamefont {{Das
  Sarma}}}, \bibinfo {author} {\bibfnamefont {X.}~\bibnamefont {Wang}}, \ and\
  \bibinfo {author} {\bibfnamefont {S.}~\bibnamefont {Yang}},\ }\href {\doibase
  10.1103/PhysRevB.83.235314} {\bibfield  {journal} {\bibinfo  {journal} {Phys.
  Rev. B}\ }\textbf {\bibinfo {volume} {83}},\ \bibinfo {pages} {235314}
  (\bibinfo {year} {2011})}\BibitemShut {NoStop}%
\bibitem [{\citenamefont {Rahman}\ \emph {et~al.}(2011)\citenamefont {Rahman},
  \citenamefont {Verduijn}, \citenamefont {Kharche}, \citenamefont
  {Lansbergen}, \citenamefont {Klimeck}, \citenamefont {Hollenberg},\ and\
  \citenamefont {Rogge}}]{Rahman:2011p195323}%
  \BibitemOpen
  \bibfield  {author} {\bibinfo {author} {\bibfnamefont {R.}~\bibnamefont
  {Rahman}}, \bibinfo {author} {\bibfnamefont {J.}~\bibnamefont {Verduijn}},
  \bibinfo {author} {\bibfnamefont {N.}~\bibnamefont {Kharche}}, \bibinfo
  {author} {\bibfnamefont {G.~P.}\ \bibnamefont {Lansbergen}}, \bibinfo
  {author} {\bibfnamefont {G.}~\bibnamefont {Klimeck}}, \bibinfo {author}
  {\bibfnamefont {L.~C.~L.}\ \bibnamefont {Hollenberg}}, \ and\ \bibinfo
  {author} {\bibfnamefont {S.}~\bibnamefont {Rogge}},\ }\href {\doibase
  10.1103/PhysRevB.83.195323} {\bibfield  {journal} {\bibinfo  {journal}
  {Physical Review B}\ }\textbf {\bibinfo {volume} {83}},\ \bibinfo {pages}
  {195323} (\bibinfo {year} {2011})}\BibitemShut {NoStop}%
\bibitem [{\citenamefont {Zhang}\ \emph {et~al.}(2013)\citenamefont {Zhang},
  \citenamefont {Luo}, \citenamefont {Saraiva}, \citenamefont {Koiller},\ and\
  \citenamefont {Zunger}}]{Zhang:2013p2396}%
  \BibitemOpen
  \bibfield  {author} {\bibinfo {author} {\bibfnamefont {L.}~\bibnamefont
  {Zhang}}, \bibinfo {author} {\bibfnamefont {J.-W.}\ \bibnamefont {Luo}},
  \bibinfo {author} {\bibfnamefont {A.}~\bibnamefont {Saraiva}}, \bibinfo
  {author} {\bibfnamefont {B.}~\bibnamefont {Koiller}}, \ and\ \bibinfo
  {author} {\bibfnamefont {A.}~\bibnamefont {Zunger}},\ }\href {\doibase
  10.1038/ncomms3396} {\bibfield  {journal} {\bibinfo  {journal} {Nature
  Communications}\ }\textbf {\bibinfo {volume} {4}},\ \bibinfo {pages} {2396}
  (\bibinfo {year} {2013})}\BibitemShut {NoStop}%
\bibitem [{\citenamefont {Yang}\ \emph {et~al.}(2012)\citenamefont {Yang},
  \citenamefont {Lim}, \citenamefont {Lai}, \citenamefont {Rossi},
  \citenamefont {Morello},\ and\ \citenamefont {Dzurak}}]{PhysRevB.86.115319}%
  \BibitemOpen
  \bibfield  {author} {\bibinfo {author} {\bibfnamefont {C.~H.}\ \bibnamefont
  {Yang}}, \bibinfo {author} {\bibfnamefont {W.~H.}\ \bibnamefont {Lim}},
  \bibinfo {author} {\bibfnamefont {N.~S.}\ \bibnamefont {Lai}}, \bibinfo
  {author} {\bibfnamefont {A.}~\bibnamefont {Rossi}}, \bibinfo {author}
  {\bibfnamefont {A.}~\bibnamefont {Morello}}, \ and\ \bibinfo {author}
  {\bibfnamefont {A.~S.}\ \bibnamefont {Dzurak}},\ }\href {\doibase
  10.1103/PhysRevB.86.115319} {\bibfield  {journal} {\bibinfo  {journal} {Phys.
  Rev. B}\ }\textbf {\bibinfo {volume} {86}},\ \bibinfo {pages} {115319}
  (\bibinfo {year} {2012})}\BibitemShut {NoStop}%
\bibitem [{\citenamefont {Wang}\ \emph {et~al.}(2011)\citenamefont {Wang},
  \citenamefont {Yang},\ and\ \citenamefont {{Das Sarma}}}]{Wang:2011p115301}%
  \BibitemOpen
  \bibfield  {author} {\bibinfo {author} {\bibfnamefont {X.}~\bibnamefont
  {Wang}}, \bibinfo {author} {\bibfnamefont {S.}~\bibnamefont {Yang}}, \ and\
  \bibinfo {author} {\bibfnamefont {S.}~\bibnamefont {{Das Sarma}}},\ }\href
  {\doibase 10.1103/PhysRevB.84.115301} {\bibfield  {journal} {\bibinfo
  {journal} {Phys. Rev. B}\ }\textbf {\bibinfo {volume} {84}},\ \bibinfo
  {pages} {115301} (\bibinfo {year} {2011})}\BibitemShut {NoStop}%
\bibitem [{\citenamefont {Pioro-Ladriere}\ \emph {et~al.}(2008)\citenamefont
  {Pioro-Ladriere}, \citenamefont {Obata}, \citenamefont {Tokura},
  \citenamefont {Shin}, \citenamefont {Kubo}, \citenamefont {Yoshida},
  \citenamefont {Taniyama},\ and\ \citenamefont
  {Tarucha}}]{PioroLadriere:2008p776}%
  \BibitemOpen
  \bibfield  {author} {\bibinfo {author} {\bibfnamefont {M.}~\bibnamefont
  {Pioro-Ladriere}}, \bibinfo {author} {\bibfnamefont {T.}~\bibnamefont
  {Obata}}, \bibinfo {author} {\bibfnamefont {Y.}~\bibnamefont {Tokura}},
  \bibinfo {author} {\bibfnamefont {Y.-S.}\ \bibnamefont {Shin}}, \bibinfo
  {author} {\bibfnamefont {T.}~\bibnamefont {Kubo}}, \bibinfo {author}
  {\bibfnamefont {K.}~\bibnamefont {Yoshida}}, \bibinfo {author} {\bibfnamefont
  {T.}~\bibnamefont {Taniyama}}, \ and\ \bibinfo {author} {\bibfnamefont
  {S.}~\bibnamefont {Tarucha}},\ }\href {\doibase 10.1038/nphys1053} {\bibfield
   {journal} {\bibinfo  {journal} {Nature Physics}\ }\textbf {\bibinfo {volume}
  {4}},\ \bibinfo {pages} {776} (\bibinfo {year} {2008})}\BibitemShut {NoStop}%
\bibitem [{\citenamefont {Assali}\ \emph {et~al.}(2011)\citenamefont {Assali},
  \citenamefont {Petrilli}, \citenamefont {Capaz}, \citenamefont {Koiller},
  \citenamefont {Hu},\ and\ \citenamefont {{Das Sarma}}}]{Assali:2011p165301}%
  \BibitemOpen
  \bibfield  {author} {\bibinfo {author} {\bibfnamefont {L.~V.~C.}\
  \bibnamefont {Assali}}, \bibinfo {author} {\bibfnamefont {H.~M.}\
  \bibnamefont {Petrilli}}, \bibinfo {author} {\bibfnamefont {R.~B.}\
  \bibnamefont {Capaz}}, \bibinfo {author} {\bibfnamefont {B.}~\bibnamefont
  {Koiller}}, \bibinfo {author} {\bibfnamefont {X.}~\bibnamefont {Hu}}, \ and\
  \bibinfo {author} {\bibfnamefont {S.}~\bibnamefont {{Das Sarma}}},\ }\href
  {\doibase 10.1103/PhysRevB.83.165301} {\bibfield  {journal} {\bibinfo
  {journal} {Phys. Rev. B}\ }\textbf {\bibinfo {volume} {83}},\ \bibinfo
  {pages} {165301} (\bibinfo {year} {2011})}\BibitemShut {NoStop}%
\bibitem [{\citenamefont {Simmons}\ \emph {et~al.}(2009)\citenamefont
  {Simmons}, \citenamefont {Thalakulam}, \citenamefont {Rosemeyer},
  \citenamefont {{van Bael}}, \citenamefont {Sackmann}, \citenamefont {Savage},
  \citenamefont {Lagally}, \citenamefont {Joynt}, \citenamefont {Friesen},
  \citenamefont {Coppersmith},\ and\ \citenamefont
  {Eriksson}}]{Simmons:2009p3234}%
  \BibitemOpen
  \bibfield  {author} {\bibinfo {author} {\bibfnamefont {C.~B.}\ \bibnamefont
  {Simmons}}, \bibinfo {author} {\bibfnamefont {M.}~\bibnamefont {Thalakulam}},
  \bibinfo {author} {\bibfnamefont {B.~M.}\ \bibnamefont {Rosemeyer}}, \bibinfo
  {author} {\bibfnamefont {B.~J.}\ \bibnamefont {{van Bael}}}, \bibinfo
  {author} {\bibfnamefont {E.~K.}\ \bibnamefont {Sackmann}}, \bibinfo {author}
  {\bibfnamefont {D.~E.}\ \bibnamefont {Savage}}, \bibinfo {author}
  {\bibfnamefont {M.~G.}\ \bibnamefont {Lagally}}, \bibinfo {author}
  {\bibfnamefont {R.}~\bibnamefont {Joynt}}, \bibinfo {author} {\bibfnamefont
  {M.}~\bibnamefont {Friesen}}, \bibinfo {author} {\bibfnamefont {S.~N.}\
  \bibnamefont {Coppersmith}}, \ and\ \bibinfo {author} {\bibfnamefont {M.~A.}\
  \bibnamefont {Eriksson}},\ }\href {\doibase 10.1021/nl9014974} {\bibfield
  {journal} {\bibinfo  {journal} {Nano Lett.}\ }\textbf {\bibinfo {volume}
  {9}},\ \bibinfo {pages} {3234} (\bibinfo {year} {2009})}\BibitemShut
  {NoStop}%
\bibitem [{\citenamefont {Fei}\ \emph {et~al.}(2015)\citenamefont {Fei},
  \citenamefont {Hung}, \citenamefont {Koh}, \citenamefont {Shim},
  \citenamefont {Coppersmith}, \citenamefont {Hu},\ and\ \citenamefont
  {Friesen}}]{PhysRevB.91.205434}%
  \BibitemOpen
  \bibfield  {author} {\bibinfo {author} {\bibfnamefont {J.}~\bibnamefont
  {Fei}}, \bibinfo {author} {\bibfnamefont {J.-T.}\ \bibnamefont {Hung}},
  \bibinfo {author} {\bibfnamefont {T.~S.}\ \bibnamefont {Koh}}, \bibinfo
  {author} {\bibfnamefont {Y.-P.}\ \bibnamefont {Shim}}, \bibinfo {author}
  {\bibfnamefont {S.~N.}\ \bibnamefont {Coppersmith}}, \bibinfo {author}
  {\bibfnamefont {X.}~\bibnamefont {Hu}}, \ and\ \bibinfo {author}
  {\bibfnamefont {M.}~\bibnamefont {Friesen}},\ }\href {\doibase
  10.1103/PhysRevB.91.205434} {\bibfield  {journal} {\bibinfo  {journal} {Phys.
  Rev. B}\ }\textbf {\bibinfo {volume} {91}},\ \bibinfo {pages} {205434}
  (\bibinfo {year} {2015})}\BibitemShut {NoStop}%
\bibitem [{\citenamefont {Kawakami}\ \emph {et~al.}(2014)\citenamefont
  {Kawakami}, \citenamefont {Scarlino}, \citenamefont {Ward}, \citenamefont
  {Braakman}, \citenamefont {Savage}, \citenamefont {Lagally}, \citenamefont
  {Friesen}, \citenamefont {Coppersmith}, \citenamefont {Eriksson},\ and\
  \citenamefont {Vandersypen}}]{Kawakami:2014p666}%
  \BibitemOpen
  \bibfield  {author} {\bibinfo {author} {\bibfnamefont {E.}~\bibnamefont
  {Kawakami}}, \bibinfo {author} {\bibfnamefont {P.}~\bibnamefont {Scarlino}},
  \bibinfo {author} {\bibfnamefont {D.~R.}\ \bibnamefont {Ward}}, \bibinfo
  {author} {\bibfnamefont {F.~R.}\ \bibnamefont {Braakman}}, \bibinfo {author}
  {\bibfnamefont {D.~E.}\ \bibnamefont {Savage}}, \bibinfo {author}
  {\bibfnamefont {M.~G.}\ \bibnamefont {Lagally}}, \bibinfo {author}
  {\bibfnamefont {M.}~\bibnamefont {Friesen}}, \bibinfo {author} {\bibfnamefont
  {S.~N.}\ \bibnamefont {Coppersmith}}, \bibinfo {author} {\bibfnamefont
  {M.~A.}\ \bibnamefont {Eriksson}}, \ and\ \bibinfo {author} {\bibfnamefont
  {L.~M.~K.}\ \bibnamefont {Vandersypen}},\ }\href {\doibase
  10.1038/nnano.2014.153} {\bibfield  {journal} {\bibinfo  {journal} {Nature
  Nanotechnology}\ }\textbf {\bibinfo {volume} {9}},\ \bibinfo {pages} {666}
  (\bibinfo {year} {2014})}\BibitemShut {NoStop}%
\bibitem [{\citenamefont {Motzoi}\ \emph {et~al.}(2009)\citenamefont {Motzoi},
  \citenamefont {Gambetta}, \citenamefont {Rebentrost},\ and\ \citenamefont
  {Wilhelm}}]{PhysRevLett.103.110501}%
  \BibitemOpen
  \bibfield  {author} {\bibinfo {author} {\bibfnamefont {F.}~\bibnamefont
  {Motzoi}}, \bibinfo {author} {\bibfnamefont {J.~M.}\ \bibnamefont
  {Gambetta}}, \bibinfo {author} {\bibfnamefont {P.}~\bibnamefont
  {Rebentrost}}, \ and\ \bibinfo {author} {\bibfnamefont {F.~K.}\ \bibnamefont
  {Wilhelm}},\ }\href {\doibase 10.1103/PhysRevLett.103.110501} {\bibfield
  {journal} {\bibinfo  {journal} {Phys. Rev. Lett.}\ }\textbf {\bibinfo
  {volume} {103}},\ \bibinfo {pages} {110501} (\bibinfo {year}
  {2009})}\BibitemShut {NoStop}%
\bibitem [{\citenamefont {Nielsen}\ and\ \citenamefont
  {Chuang}(2000)}]{Nielsen:2000}%
  \BibitemOpen
  \bibfield  {author} {\bibinfo {author} {\bibfnamefont {M.~A.}\ \bibnamefont
  {Nielsen}}\ and\ \bibinfo {author} {\bibfnamefont {I.~L.}\ \bibnamefont
  {Chuang}},\ }\href@noop {} {\emph {\bibinfo {title} {Quantum Computation and
  Quantum Information}}}\ (\bibinfo  {publisher} {Cambridge University Press},\
  \bibinfo {address} {Cambridge},\ \bibinfo {year} {2000})\BibitemShut
  {NoStop}%
\bibitem [{\citenamefont {Barrett}\ and\ \citenamefont
  {Barnes}(2002)}]{Barrett:2002p125318}%
  \BibitemOpen
  \bibfield  {author} {\bibinfo {author} {\bibfnamefont {S.~D.}\ \bibnamefont
  {Barrett}}\ and\ \bibinfo {author} {\bibfnamefont {C.~H.~W.}\ \bibnamefont
  {Barnes}},\ }\href {\doibase 10.1103/PhysRevB.66.125318} {\bibfield
  {journal} {\bibinfo  {journal} {Phys. Rev. B}\ }\textbf {\bibinfo {volume}
  {66}},\ \bibinfo {pages} {125318} (\bibinfo {year} {2002})}\BibitemShut
  {NoStop}%
\bibitem [{\citenamefont {Dial}\ \emph {et~al.}(2013)\citenamefont {Dial},
  \citenamefont {Shulman}, \citenamefont {Harvey}, \citenamefont {Bluhm},
  \citenamefont {Umansky},\ and\ \citenamefont
  {Yacoby}}]{PhysRevLett.110.146804}%
  \BibitemOpen
  \bibfield  {author} {\bibinfo {author} {\bibfnamefont {O.~E.}\ \bibnamefont
  {Dial}}, \bibinfo {author} {\bibfnamefont {M.~D.}\ \bibnamefont {Shulman}},
  \bibinfo {author} {\bibfnamefont {S.~P.}\ \bibnamefont {Harvey}}, \bibinfo
  {author} {\bibfnamefont {H.}~\bibnamefont {Bluhm}}, \bibinfo {author}
  {\bibfnamefont {V.}~\bibnamefont {Umansky}}, \ and\ \bibinfo {author}
  {\bibfnamefont {A.}~\bibnamefont {Yacoby}},\ }\href {\doibase
  10.1103/PhysRevLett.110.146804} {\bibfield  {journal} {\bibinfo  {journal}
  {Phys. Rev. Lett.}\ }\textbf {\bibinfo {volume} {110}},\ \bibinfo {pages}
  {146804} (\bibinfo {year} {2013})}\BibitemShut {NoStop}%
\bibitem [{\citenamefont {Taylor}\ \emph {et~al.}(2007)\citenamefont {Taylor},
  \citenamefont {Petta}, \citenamefont {Johnson}, \citenamefont {Yacoby},
  \citenamefont {Marcus},\ and\ \citenamefont {Lukin}}]{PhysRevB.76.035315}%
  \BibitemOpen
  \bibfield  {author} {\bibinfo {author} {\bibfnamefont {J.~M.}\ \bibnamefont
  {Taylor}}, \bibinfo {author} {\bibfnamefont {J.~R.}\ \bibnamefont {Petta}},
  \bibinfo {author} {\bibfnamefont {A.~C.}\ \bibnamefont {Johnson}}, \bibinfo
  {author} {\bibfnamefont {A.}~\bibnamefont {Yacoby}}, \bibinfo {author}
  {\bibfnamefont {C.~M.}\ \bibnamefont {Marcus}}, \ and\ \bibinfo {author}
  {\bibfnamefont {M.~D.}\ \bibnamefont {Lukin}},\ }\href {\doibase
  10.1103/PhysRevB.76.035315} {\bibfield  {journal} {\bibinfo  {journal} {Phys.
  Rev. B}\ }\textbf {\bibinfo {volume} {76}},\ \bibinfo {pages} {035315}
  (\bibinfo {year} {2007})}\BibitemShut {NoStop}%
\bibitem [{\citenamefont {Schrieffer}\ and\ \citenamefont
  {Wolf}(1966)}]{Schrieffer:1966p491}%
  \BibitemOpen
  \bibfield  {author} {\bibinfo {author} {\bibfnamefont {J.~R.}\ \bibnamefont
  {Schrieffer}}\ and\ \bibinfo {author} {\bibfnamefont {P.~A.}\ \bibnamefont
  {Wolf}},\ }\href {\doibase 10.1103/PhysRev.149.491} {\bibfield  {journal}
  {\bibinfo  {journal} {Physical Review}\ }\textbf {\bibinfo {volume} {149}},\
  \bibinfo {pages} {491} (\bibinfo {year} {1966})}\BibitemShut {NoStop}%
\bibitem [{\citenamefont {Gros}\ \emph {et~al.}(1987)\citenamefont {Gros},
  \citenamefont {Joynt},\ and\ \citenamefont {Rice}}]{Gros:1987p381}%
  \BibitemOpen
  \bibfield  {author} {\bibinfo {author} {\bibfnamefont {C.}~\bibnamefont
  {Gros}}, \bibinfo {author} {\bibfnamefont {R.}~\bibnamefont {Joynt}}, \ and\
  \bibinfo {author} {\bibfnamefont {T.~M.}\ \bibnamefont {Rice}},\ }\href
  {\doibase 10.1103/PhysRevB.36.381} {\bibfield  {journal} {\bibinfo  {journal}
  {Phys. Rev. B}\ }\textbf {\bibinfo {volume} {36}},\ \bibinfo {pages} {381}
  (\bibinfo {year} {1987})}\BibitemShut {NoStop}%
\bibitem [{\citenamefont {MacDonald}\ \emph {et~al.}(1988)\citenamefont
  {MacDonald}, \citenamefont {Girvin},\ and\ \citenamefont
  {Yoshioka}}]{MacDonald:1988p9753}%
  \BibitemOpen
  \bibfield  {author} {\bibinfo {author} {\bibfnamefont {A.~H.}\ \bibnamefont
  {MacDonald}}, \bibinfo {author} {\bibfnamefont {S.~M.}\ \bibnamefont
  {Girvin}}, \ and\ \bibinfo {author} {\bibfnamefont {D.}~\bibnamefont
  {Yoshioka}},\ }\href {\doibase 10.1103/PhysRevB.37.9753} {\bibfield
  {journal} {\bibinfo  {journal} {Phys. Rev. B}\ }\textbf {\bibinfo {volume}
  {37}},\ \bibinfo {pages} {9753} (\bibinfo {year} {1988})}\BibitemShut
  {NoStop}%
\bibitem [{\citenamefont {Vutha}(2010)}]{LZ-ref}%
  \BibitemOpen
  \bibfield  {author} {\bibinfo {author} {\bibfnamefont {A.~C.}\ \bibnamefont
  {Vutha}},\ }\href {http://stacks.iop.org/0143-0807/31/i=2/a=016} {\bibfield
  {journal} {\bibinfo  {journal} {European Journal of Physics}\ }\textbf
  {\bibinfo {volume} {31}},\ \bibinfo {pages} {389} (\bibinfo {year}
  {2010})}\BibitemShut {NoStop}%
\bibitem [{\citenamefont {Shi}\ \emph {et~al.}(2013)\citenamefont {Shi},
  \citenamefont {Simmons}, \citenamefont {Ward}, \citenamefont {Prance},
  \citenamefont {Wu}, \citenamefont {Koh}, \citenamefont {Gamble},
  \citenamefont {Savage}, \citenamefont {Lagally}, \citenamefont {Friesen},
  \citenamefont {Coppersmith},\ and\ \citenamefont {Eriksson}}]{Shi:2013p3020}%
  \BibitemOpen
  \bibfield  {author} {\bibinfo {author} {\bibfnamefont {Z.}~\bibnamefont
  {Shi}}, \bibinfo {author} {\bibfnamefont {C.~B.}\ \bibnamefont {Simmons}},
  \bibinfo {author} {\bibfnamefont {D.~R.}\ \bibnamefont {Ward}}, \bibinfo
  {author} {\bibfnamefont {J.~R.}\ \bibnamefont {Prance}}, \bibinfo {author}
  {\bibfnamefont {X.}~\bibnamefont {Wu}}, \bibinfo {author} {\bibfnamefont
  {T.~S.}\ \bibnamefont {Koh}}, \bibinfo {author} {\bibfnamefont {J.~K.}\
  \bibnamefont {Gamble}}, \bibinfo {author} {\bibfnamefont {D.~E.}\
  \bibnamefont {Savage}}, \bibinfo {author} {\bibfnamefont {M.~G.}\
  \bibnamefont {Lagally}}, \bibinfo {author} {\bibfnamefont {M.}~\bibnamefont
  {Friesen}}, \bibinfo {author} {\bibfnamefont {S.~N.}\ \bibnamefont
  {Coppersmith}}, \ and\ \bibinfo {author} {\bibfnamefont {M.~A.}\ \bibnamefont
  {Eriksson}},\ }\href {\doibase doi:10.1038/ncomms4020} {\bibfield  {journal}
  {\bibinfo  {journal} {Nature Communications}\ }\textbf {\bibinfo {volume}
  {5}},\ \bibinfo {pages} {3020} (\bibinfo {year} {2013})}\BibitemShut
  {NoStop}%
\bibitem [{\citenamefont {Eng}\ \emph {et~al.}(2015)\citenamefont {Eng},
  \citenamefont {Ladd}, \citenamefont {Smith}, \citenamefont {Borselli},
  \citenamefont {Kiselev}, \citenamefont {Fong}, \citenamefont {Holabird},
  \citenamefont {Hazard}, \citenamefont {Huang}, \citenamefont {Deelman},
  \citenamefont {Milosavljevic}, \citenamefont {Schmitz}, \citenamefont {Ross},
  \citenamefont {Gyure},\ and\ \citenamefont {Hunter}}]{Eng:2015pe1500214}%
  \BibitemOpen
  \bibfield  {author} {\bibinfo {author} {\bibfnamefont {K.}~\bibnamefont
  {Eng}}, \bibinfo {author} {\bibfnamefont {T.~D.}\ \bibnamefont {Ladd}},
  \bibinfo {author} {\bibfnamefont {A.}~\bibnamefont {Smith}}, \bibinfo
  {author} {\bibfnamefont {M.~G.}\ \bibnamefont {Borselli}}, \bibinfo {author}
  {\bibfnamefont {A.~A.}\ \bibnamefont {Kiselev}}, \bibinfo {author}
  {\bibfnamefont {B.~H.}\ \bibnamefont {Fong}}, \bibinfo {author}
  {\bibfnamefont {K.~S.}\ \bibnamefont {Holabird}}, \bibinfo {author}
  {\bibfnamefont {T.~M.}\ \bibnamefont {Hazard}}, \bibinfo {author}
  {\bibfnamefont {B.}~\bibnamefont {Huang}}, \bibinfo {author} {\bibfnamefont
  {P.~W.}\ \bibnamefont {Deelman}}, \bibinfo {author} {\bibfnamefont
  {I.}~\bibnamefont {Milosavljevic}}, \bibinfo {author} {\bibfnamefont {A.~E.}\
  \bibnamefont {Schmitz}}, \bibinfo {author} {\bibfnamefont {R.~S.}\
  \bibnamefont {Ross}}, \bibinfo {author} {\bibfnamefont {M.~F.}\ \bibnamefont
  {Gyure}}, \ and\ \bibinfo {author} {\bibfnamefont {A.~T.}\ \bibnamefont
  {Hunter}},\ }\href {\doibase 10.1126/sciadv.1500214} {\bibfield  {journal}
  {\bibinfo  {journal} {Science Advances}\ }\textbf {\bibinfo {volume} {1}},\
  \bibinfo {pages} {e1500214} (\bibinfo {year} {2015})}\BibitemShut {NoStop}%
\bibitem [{\citenamefont {H\"uttel}\ \emph {et~al.}(2005)\citenamefont
  {H\"uttel}, \citenamefont {Ludwig}, \citenamefont {Lorenz}, \citenamefont
  {Eberl},\ and\ \citenamefont {Kotthaus}}]{PhysRevB.72.081310}%
  \BibitemOpen
  \bibfield  {author} {\bibinfo {author} {\bibfnamefont {A.~K.}\ \bibnamefont
  {H\"uttel}}, \bibinfo {author} {\bibfnamefont {S.}~\bibnamefont {Ludwig}},
  \bibinfo {author} {\bibfnamefont {H.}~\bibnamefont {Lorenz}}, \bibinfo
  {author} {\bibfnamefont {K.}~\bibnamefont {Eberl}}, \ and\ \bibinfo {author}
  {\bibfnamefont {J.~P.}\ \bibnamefont {Kotthaus}},\ }\href {\doibase
  10.1103/PhysRevB.72.081310} {\bibfield  {journal} {\bibinfo  {journal} {Phys.
  Rev. B}\ }\textbf {\bibinfo {volume} {72}},\ \bibinfo {pages} {081310}
  (\bibinfo {year} {2005})}\BibitemShut {NoStop}%
\bibitem [{\citenamefont {Thalakulam}\ \emph {et~al.}(2010)\citenamefont
  {Thalakulam}, \citenamefont {Simmons}, \citenamefont {Rosemeyer},
  \citenamefont {Savage}, \citenamefont {Lagally}, \citenamefont {Friesen},
  \citenamefont {Coppersmith},\ and\ \citenamefont
  {Eriksson}}]{Thalakulam:2010p183104}%
  \BibitemOpen
  \bibfield  {author} {\bibinfo {author} {\bibfnamefont {M.}~\bibnamefont
  {Thalakulam}}, \bibinfo {author} {\bibfnamefont {C.~B.}\ \bibnamefont
  {Simmons}}, \bibinfo {author} {\bibfnamefont {B.~M.}\ \bibnamefont
  {Rosemeyer}}, \bibinfo {author} {\bibfnamefont {D.~E.}\ \bibnamefont
  {Savage}}, \bibinfo {author} {\bibfnamefont {M.~G.}\ \bibnamefont {Lagally}},
  \bibinfo {author} {\bibfnamefont {M.}~\bibnamefont {Friesen}}, \bibinfo
  {author} {\bibfnamefont {S.~N.}\ \bibnamefont {Coppersmith}}, \ and\ \bibinfo
  {author} {\bibfnamefont {M.~A.}\ \bibnamefont {Eriksson}},\ }\href {\doibase
  10.1063/1.3425892} {\bibfield  {journal} {\bibinfo  {journal} {Appl. Phys.
  Lett.}\ }\textbf {\bibinfo {volume} {96}},\ \bibinfo {pages} {183104}
  (\bibinfo {year} {2010})}\BibitemShut {NoStop}%
\end{thebibliography}%

\end{document}